\documentclass[sigconf,authorversion,nonacm]{acmart}

\usepackage{float}
\usepackage[normalem]{ulem} %
\usepackage[multiple]{footmisc}

\AtBeginDocument{%
  \providecommand\BibTeX{{%
    \normalfont B\kern-0.5em{\scshape i\kern-0.25em b}\kern-0.8em\TeX}}}

\setcopyright{acmcopyright}
\copyrightyear{2023}
\acmYear{2023}
\acmDOI{XXXXXXX.XXXXXXX}

\begin{document}

\title[The Semantic Reader Project]{The Semantic Reader Project: Augmenting  Scholarly Documents through AI-Powered Interactive Reading Interfaces}
\titlenote{[\href{https://www.semanticscholar.org/reader/67a5bacb00651dbe0dd9ef2a563fe64b19b2c6a8}{\textcolor{blue}{\dotuline{Click to open in the Semantic Reader}}}]\hspace{3mm}[\href{https://ai2-s2-research-public.s3.amazonaws.com/semantic-reader/the-semantic-reader-project-augmenting-scholarly-documents-through-ai-powered-interactive-reading-interfaces.pdf}{\textcolor{blue}{\dotuline{Download the version with alt-text}}}]}

\author{
Kyle Lo$^{\alpha}$\hspace{0.1em} 
Joseph Chee Chang$^{\alpha}$\hspace{0.1em}
Andrew Head$^{\psi}$\hspace{0.1em}  
Jonathan Bragg$^{\alpha}$\hspace{0.1em}  
Amy X. Zhang$^{\omega}$\hspace{0.1em}
Cassidy Trier$^{\alpha}$ \\ \vspace{0.5em}
Chloe Anastasiades$^{\alpha}$\hspace{0.1em} 
Tal August$^\alpha$\hspace{0.1em} 
Russell Authur$^{\alpha}$\hspace{0.1em}
Danielle Bragg$^{\rho}$\hspace{0.1em} 
Erin Bransom$^\alpha$\hspace{0.1em} \\
Isabel Cachola$^{\iota}$\hspace{0.1em}
Stefan Candra$^\alpha$\hspace{0.1em} 
Yoganand Chandrasekhar$^\alpha$\hspace{0.1em} 
Yen-Sung Chen$^{\alpha}$\hspace{0.1em}   \\
Evie Yu-Yen Cheng$^{\alpha}$\hspace{0.1em}
Yvonne Chou$^{\alpha}$\hspace{0.1em}
Doug Downey$^{\alpha}$\hspace{0.1em}  
Rob Evans$^{\alpha}$\hspace{0.1em} 
Raymond Fok$^{\omega}$\hspace{0.1em}  \\
Fangzhou Hu$^{\alpha}$\hspace{0.1em} 
Regan Huff$^{\alpha}$\hspace{0.1em} 
Dongyeop Kang$^{\upsilon}$\hspace{0.1em} 
Tae Soo Kim$^{\kappa}$\hspace{0.1em} 
Rodney Kinney$^{\alpha}$\hspace{0.1em}  \\
Aniket Kittur$^{\chi}$\hspace{0.1em} 
Hyeonsu Kang$^{\chi}$\hspace{0.1em} 
Egor Klevak$^{\alpha}$\hspace{0.1em} 
Bailey Kuehl$^{\alpha}$\hspace{0.1em} 
Michael Langan$^{\alpha}$\hspace{0.1em} \\
Matt Latzke$^{\alpha}$\hspace{0.1em} 
Jaron Lochner$^{\alpha}$\hspace{0.1em} 
Kelsey MacMillan$^{\alpha}$\hspace{0.1em}
Eric Marsh$^{\alpha}$\hspace{0.1em} 
Tyler Murray$^{\alpha}$\hspace{0.1em}   \\
Aakanksha Naik$^{\alpha}$\hspace{0.1em}
Ngoc-Uyen Nguyen$^{\alpha}$\hspace{0.1em} 
Srishti Palani$^{\sigma}$\hspace{0.1em}  
Soya Park$^{\tau}$\hspace{0.1em} 
Caroline Paulic$^{\alpha}$\hspace{0.1em} \\
Napol Rachatasumrit$^{\chi}$\hspace{0.1em}
Smita Rao$^{\alpha}$ \hspace{0.1em} 
Paul Sayre$^{\alpha}$ \hspace{0.1em}
Zejiang Shen$^{\tau}$\hspace{0.1em} 
Pao Siangliulue$^{\alpha}$\hspace{0.1em}  \\
Luca Soldaini$^{\alpha}$\hspace{0.1em} 
Huy Tran$^{\alpha}$\hspace{0.1em}
Madeleine van Zuylen$^{\alpha}$\hspace{0.1em}
Lucy Lu Wang$^{\omega}$\hspace{0.1em}  \\
Christopher Wilhelm$^{\alpha}$\hspace{0.1em}
Caroline Wu$^{\alpha}$\hspace{0.1em}
Jiangjiang Yang$^{\alpha}$\hspace{0.1em} 
Angele Zamarron$^{\alpha}$\hspace{0.1em}\\ \vspace{0.5em} 
Marti A. Hearst$^{\beta}$\hspace{0.1em} 
Daniel S. Weld$^{\alpha}$\vspace{0.5em}
}
  \affiliation{
  $^{\alpha}$Allen Institute for AI  \hspace{0.1em} 
  $^{\omega}$University of Washington \hspace{0.1em} 
  $^{\beta}$University of California, Berkeley \\
  $^{\psi}$University of Pennsylvania \hspace{0.1em} 
  $^{\chi}$Carnegie Mellon University \hspace{0.1em} 
  $^{\tau}$Massachusetts Institute of Technology \hspace{0.1em} 
  $^{\kappa}$KAIST \hspace{0.1em} \\
  $^{\iota}$Johns Hopkins University \hspace{0.1em}
  $^{\upsilon}$University of Minnesota \hspace{0.1em} 
  $^{\sigma}$University of California, San Diego \hspace{0.1em} 
  $^{\rho}$Microsoft Research \vspace{0.5em}
  \country{}
  }
  \email{{kylel, josephc, jbragg, cassidyt, danw}@allenai.org,
  head@seas.penn.edu, axz@cs.uw.edu, hearst@berkeley.edu}

\renewcommand{\shortauthors}{Lo and Chang, et al.}

\newcommand{\bug}
    {\mbox{\rule{2mm}{2mm}}}
\newcommand{\Bug}[1]
    {\bug \footnote{BUG: {#1}}}
\newcommand{\andrew}[1]{\textcolor{purple}{\textit{Andrew}: #1}}
\newcommand{\joseph}[1]{\textcolor{blue}{\textit{Joseph}: #1}}
\newcommand{\kyle}[1]{\textcolor{orange}{\textit{Kyle}: #1}}
\newcommand{\jonathan}[1]{\textcolor{green}{\textit{Jonathan}: #1}}
\newcommand{\amy}[1]{\textcolor{brown}{\textit{Amy}: #1}}
\newcommand{\marti}[1]{\textcolor{purple}{\textit{Marti}: #1}}
\newcommand{\DW}[1]{\bug \footnote{\textcolor{blue}{Dan: #1}}}
\newcommand{\dan}[1]{\textcolor{blue}{#1}}
\newcommand{\cassidy}[1]{\textcolor{orange}{\textit{Cassidy}: #1}}
\newcommand{\tal}[1]{\textcolor{teal}{\textit{Tal}: #1}}
\newcommand{\ray}[1]{\textcolor{red}{\textit{Ray}: #1}}
\newcommand{\hyeonsu}[1]{\textcolor{teal}{\textit{Hyeonsu}: #1}}
\newcommand{\pao}[1]{\textcolor{pink}{\textit{Pao}: #1}}

\makeatletter
\newcommand\footnoteref[1]{\protected@xdef\@thefnmark{\ref{#1}}\@footnotemark}
\makeatother

\newcommand{\eg}{\textit{e.g., }}
\newcommand{\cf}{\textit{cf. }}

\newcommand{\User}{Reader}
\newcommand{\Users}{{\User}s}
\newcommand{\user}{reader}
\newcommand{\users}{{\user}s}

\newcommand{\System}{Reading interface}
\newcommand{\Systems}{{\System}s}
\newcommand{\system}{reading interface}
\newcommand{\systems}{{\system}s}

\newcommand{\Paper}{Research paper}
\newcommand{\Papers}{{\Paper}s}
\newcommand{\paper}{research paper}
\newcommand{\papers}{{\paper}s}

\begin{abstract}
Scholarly publications are key to the transfer of knowledge from scholars to others. However, research papers are information-dense, and as the volume of the scientific literature grows, the need for new technology to support the reading process grows.
In contrast to the process of {\em finding} papers, which has been transformed by Internet technology, the experience of {\em reading} {\papers} has changed little in decades. 
The PDF format for sharing papers is widely used due to its portability, but it has significant downsides including: static content, poor accessibility for low-vision readers, and difficulty reading on mobile devices.
This paper explores the question ``Can recent advances in AI and HCI power intelligent, interactive, and accessible reading interfaces---even for legacy PDFs?'' 
We describe the Semantic Reader Project, a collaborative effort across multiple institutions to explore automatic creation of dynamic reading interfaces for research papers. Through this project, we've developed ten research prototype interfaces and conducted usability studies with 300+ participants and real-world users showing improved reading experiences for scholars. We've also released a production research paper reader that will incorporate novel features as they mature. We structure this paper around challenges scholars and the public face when reading {\papers}---discovery, efficiency, comprehension, synthesis, and accessibility---and present an overview of our progress and remaining open challenges.
\end{abstract}

\maketitle

\section{Introduction}

The exponential growth of scientific publication~\cite{Bornmann2020GrowthRO,brainard2020scientists} and increasing interdisciplinary nature of scientific progress \cite{van2015interdisciplinary, okamura2019interdisciplinarity} makes it increasingly hard for scholars to keep up with the latest developments. 
Academic search engines, such as Google Scholar and Semantic Scholar
help scholars discover research papers. 
Automated summarization for research papers~\cite{Cachola2020TLDRES} helps scholars triage between {\papers}.  
But when it comes to actually {\em reading} {\papers}, the process, based on a static PDF format, has remained largely unchanged for many decades.
This is a problem because digesting technical research papers is difficult~\cite{Bazerman1985-la,bem1995writing}.%

In contrast, interactive and personalized documents have seen significant adoption in domains outside of academic research. For example, news websites such as the \textit{New York Times} often present interactive articles with explorable visualizations that allow readers to understand complex data in a personalized way. E-readers, such as the Kindle, provide in-situ context to help readers better comprehend complex documents, showing inline term definitions and tracking occurrence of characters in a long novel. 
While prior work has focused on authoring support tools~\cite{conlen2021idyll,Latif2021KoriIS,Conlen2022FidyllAC} that can reduce effort in creating interactive scientific documents~\cite{Hohman2020CommunicatingWI,head2022math}, they have not seen widespread adoption due to a lack of incentive structure \cite{Distill_Editorial_Team2021-ix}. 
Furthermore, millions of {\papers} are locked in the rigid and static PDF format, whose low-level syntax makes it extremely difficult for systems to access semantic content, augment interactivity, or even provide basic reading functionality for assistive tools like screen readers~\cite{bigham-uninteresting-tour}.

\begin{figure*}[ht!]
    \centering
    \includegraphics[width=0.8\textwidth]{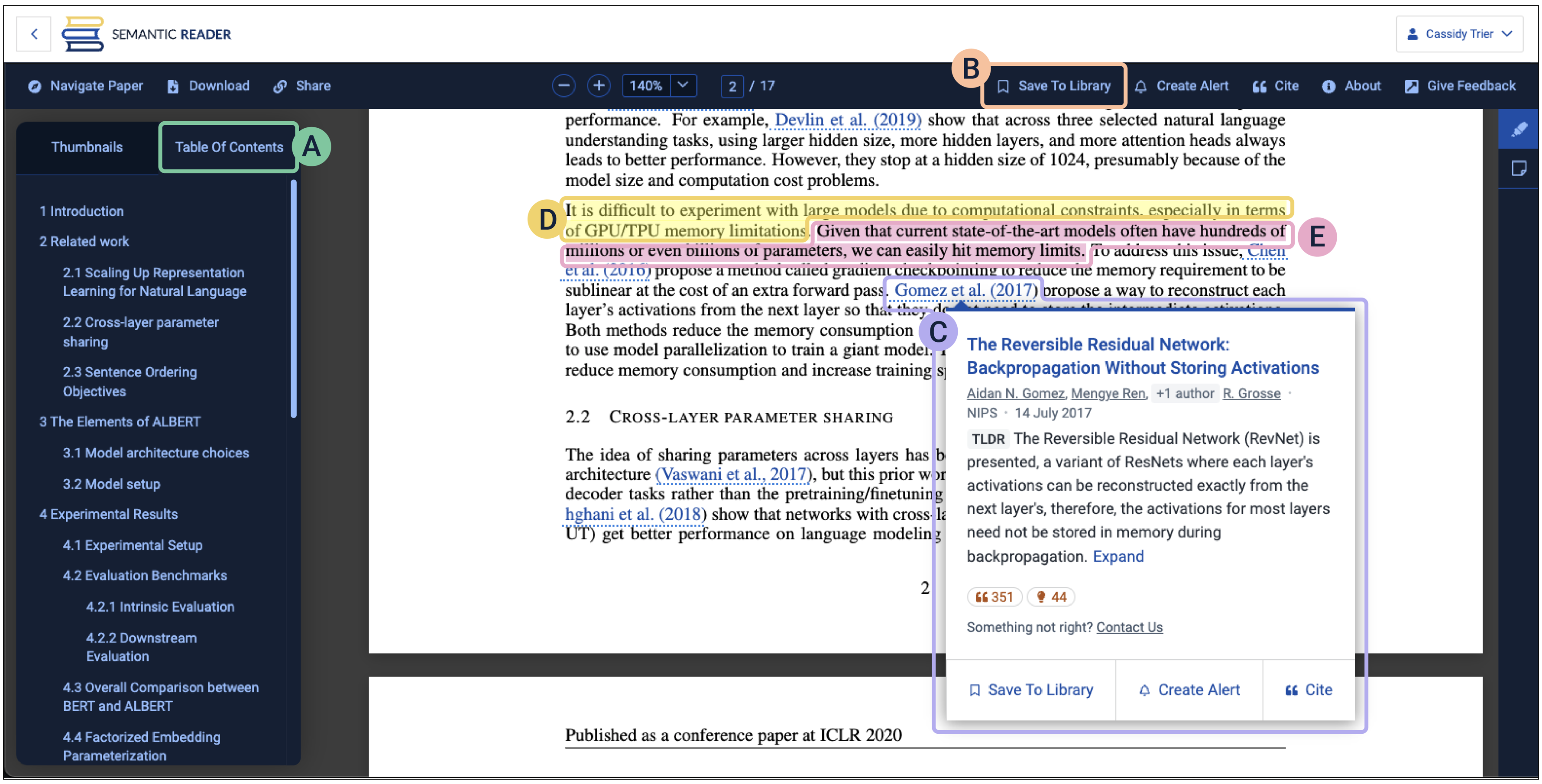}
    \caption{The Semantic Reader Project consists of research, product, and open science resources. The Semantic Reader product\footref{product} is a free interactive interface for research papers. It supports standard reading features (e.g., (A) table of contents), integration with Semantic Scholar (e.g., (B) save to library), useful augmentations atop the existing PDF (e.g., (C) in-situ Paper Cards when clicking inline citations), and integration with third-party features (e.g. (D) Hypothes.is\footref{hypothesis} for user highlights).
    We continues to integrate research features into this product as they mature (e.g., (E) Scim automated highlights \S\ref{sec:scim}).
    } 

    \label{fig:product}
\end{figure*}

Fortunately, recent work on layout-aware document parsing~\cite{Xu2019LayoutLMPO,Huang2022LayoutLMv3PF,shen-etal-2022-vila}
and large language models~\cite{Beltagy2019SciBERTAP,Raffel2019ExploringTL,gpt3-brown-2020}
show promise for accessing the content of PDF documents, and building systems that can better understand their semantics. 
This raises an exciting challenge: \emph{Can we create intelligent, interactive, and accessible reading interfaces for research papers, even atop existing PDFs?}

To explore this question, we present the \textbf{Semantic Reader Project}, a broad collaborative effort across multiple non-profit, industry, and academic institutions to create interactive, intelligent reading interfaces for research papers. 
This project consists of three pillars: research, product, and open science resources.
On the research front, the Semantic Reader Project combines AI and HCI research to design novel, AI-powered interactive reading interfaces that address a variety of user challenges faced by today's scholars. We developed research prototypes and conducted usability studies that clarify their benefits.
On the product front, we are developing the Semantic Reader~(Figure~\ref{fig:product}),\footnote{\label{product}Semantic Reader: \url{ https://www.semanticscholar.org/product/semantic-reader}} 
a freely available reading interface that integrates features from research prototypes as they mature.\footnote{Available for over 369K papers as of February 2023.}
Finally, we are developing and releasing open science resources that drive both the research and the product. These resources together open-source software,\footnote{\label{library}For UI development: \url{https://github.com/allenai/pdf-component-library}}\footnote{\label{papermage}For processing PDFs: \url{https://github.com/allenai/papermage}} AI models \cite{cohan2020specter,Cachola2020TLDRES,shen-etal-2022-vila,kang-etal-2020-document}, and open datasets \cite{kinney2023semantic,lo-etal-2020-s2orc} to support continued work in this area.%

In this paper, we focus on summarizing our efforts under the \emph{research pillar} of the Semantic Reader Project. We structure our discussion around five broad challenges faced by readers of research papers:
\addtocounter{footnote}{1} %
\footnotetext{\label{hypothesis}Hypothes.is: https://web.hypothes.is\url{https://web.hypothes.is}}

\begin{itemize}
  \setlength\itemsep{0.25em}

    \item \textbf{Discovery:} Following paper citations is one of the main strategies that scholars employ to discover additional relevant papers, but keeping track of the large numbers of citations can be overwhelming. In \S\ref{sec:citation-discovery}, we explore ways to visually augment research papers to help readers prioritize their paper exploration during literature reviews.
    \item \textbf{Efficiency:} The exponential growth of publication makes it difficult for scholars to keep up-to-date with the literature---scholars need to skim and read many papers while making sure they capture enough details in each. In \S\ref{sec:guiding}, we explore how support for non-linear reading can help readers consume research papers more efficiently.
    \item \textbf{Comprehension:} Research papers can be dense and contain terms that are unfamiliar either because the author newly introduces them or assumes readers have prerequisite domain knowledge. In \S\ref{sec:in-situ-explanations}, we explore how providing in-situ definitions and summaries can benefit readers especially when reading outside of their domains.
    \item \textbf{Synthesis:} The sensemaking \cite{Russell_Sensemaking_1993} process of synthesizing knowledge scattered across multiple papers
    is effortful but important. It allows scholars to make connections between prior work and identify opportunities for future research. In \S\ref{sec:bootstrap}, we explore how to help readers collect information from and make sense of many papers to gain better understanding of broad research topics.%
    \item \textbf{Accessibility:} Static PDFs are an ill-suited format for many reading interfaces. For example, PDFs are notoriously incompatible with screen readers, and represent a significant barrier for blind and low vision readers~\cite{bigham-uninteresting-tour}. Furthermore, an increasing number of scholars access content on mobile devices, on which PDFs of papers are difficult to read. In \S\ref{sec:accessibility}, we explore methods for converting legacy papers to more accessible representations.

\end{itemize}
\noindent Specifically, we present ten research prototypes developed in the Semantic Reader Project---CiteSee~\cite{Chang2022CiteSee}, CiteRead~\cite{Rachatasumrit2022CiteReadIL}, 
Scim~\cite{Fok2023Scim}, 
Ocean~\cite{park-cscw22}, ScholarPhi~\cite{Head2021AugmentingSP}, Paper Plain~\cite{August2022PaperPM}, Papeo~\cite{papeo}, Threddy~\cite{Kang2022Threddy}, Relatedly~\cite{relatedly}, and SciA11y~\cite{wang-2021-scia11y,paper2html}---and explain how they address these reading challenges. 
We conclude by discussing ongoing research opportunities in both AI and HCI for developing the future of scholarly reading interfaces. We provide pointers to our production reading interface and associated open resources to invite the broader research community to join our effort.

\section{Unlocking Citations for Discovery}
\label{sec:citation-discovery}

Scholars use many methods to discover relevant {\papers} to read, including search engines, word of mouth, and browsing familiar venues. However, once they find one {\paper}, it's especially common for scholars to use its references and citations  to further expand their knowledge of a research area. This behavior, sometimes referred to as \emph{forward/backward chaining} or \emph{footnote chasing}, is ubiquitous and has been observed across many scholarly disciplines \cite{palmer2009scholarly}.
Supporting this, one popular feature in the Semantic Reader\footnoteref{product} is in-situ Paper Cards that pop up when {\users} click on an inline citation, dramatically reducing the interaction cost caused by 
jumping back-and-forth between inline citations and their corresponding references at the end of a {\paper} (Figure~\ref{fig:product}).
Despite this affordance, during literature reviews, {\users} may still be overwhelmed trying to make sense of the tens to hundreds of inline citations in each paper \cite{Denney2013HowTW,Peroni2015SettingOB,Chang2022CiteSee}.
Conversely, when reading a given paper, a reader cannot see  relevant follow-on {\papers} that cited the current paper.
Here we discuss how interactive {\systems} can help scholars more effectively explore citations to important relevant work in both directions with two systems called CiteSee \cite{Chang2022CiteSee} and CiteRead \cite{Rachatasumrit2022CiteReadIL}.

\subsection{Augmenting Citations with CiteSee}
\label{sec:citesee}

\begin{figure}[t]
    \centering
    \includegraphics[width=1\columnwidth]{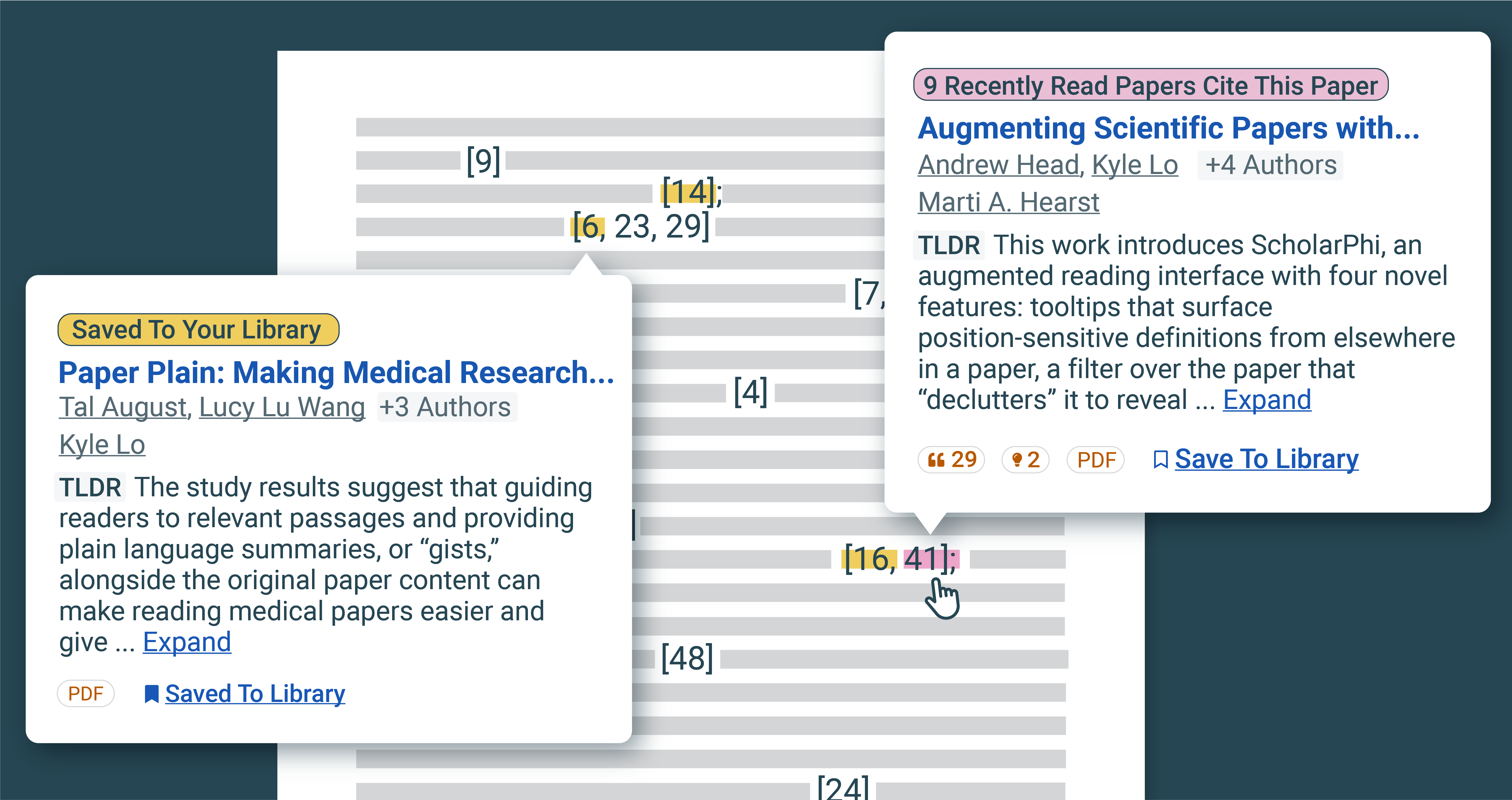}
    \caption{
CiteSee \cite{Chang2022CiteSee} highlights citations to familiar papers (e.g., recently read or saved in their libraries) 
as well as unfamiliar papers to help readers avoid overlooking important citations when conducting literature reviews.
    Clicking on \emph{Expand} surfaces additional context, such as citing sentences from recently read papers.
    }
    \label{fig:citation-discovery}
\end{figure}

While most prior work on supporting {\paper} discovery has focused on developing bespoke interfaces of recommender systems or visualizations based on paper contents \cite{Sugiyama2010ScholarlyPR,Philip2014ApplicationOC}, the citation graph \cite{Huang2002AGR,Gori2006ResearchPR,Xia2016ScientificAR,Mackinlay1995AnOU,chau2011apolo,He2019PaperPolesFA,Ponsard2016PaperQuestAV}, or a combination of the two \cite{Wang2011CollaborativeTM,cohan2020specter}, {\paper} discovery via inline citations in a reading interface is important but under-explored. 
One study estimates that reading and exploring inline citations accounts for around one in five  {\paper} discoveries during active research \cite{King2009ScholarlyJI}. However, while all inline citations are relevant to the current {\paper}, it is likely that some are more relevant to the current {\user} than others.
For example, a {\user} reading papers about \emph{aspect extraction of online product reviews} to learn more about \emph{natural language processing techniques} would be less interested in citations to {\papers} around \emph{e-commerce and marketing}.
In addition,  citations to the same {\papers} often have different surface forms across papers (i.e., reference numbers), making it all the more difficult for {\users} to keep track of all the inline citations they should explore or have already explored during literature reviews.

To address this, CiteSee provides a personalized {\paper} reading experience by automatically identifying and resolving inline citations in PDFs to research paper entities in our academic graph \cite{kinney2023semantic}, and visually augmenting inline citations based on their connections to the current {\user}.
First, CiteSee leverages a {\user}'s reading behavior and history as a way to capture their short-term and fluid interests during literature reviews. Using this signal, CiteSee scores and highlights inline citations to help the {\user} triage them and discover prior work that are likely relevant to their literature review topics (Figure~\ref{fig:citation-discovery}). Second, CiteSee leverages {\papers} saved in the {\user}'s Semantic Scholar paper library and the {\user}'s publication record \cite{kinney2023semantic} to understand their longer-term research interests. Using this signal, CiteSee changes the colors of the inline citations to familiar papers so that the {\user} can both better contextualize the current paper and keep track of citations to papers they have already explored.
In addition, CiteSee also helps {\users} better make sense of the cited papers by showing how they connect to a {\user}'s previous activities; for example, showing which library folders they were saved under or the citing sentences from a familiar {\paper} (Figure~\ref{fig:citation-discovery}).
Based on lab and field studies, CiteSee showed promise that providing visual augmentation and personalized context around inline citations in an interactive reading environment can allow {\users} to more effectively discover relevant prior work and keep track of their exploration during real-world literature review tasks.

\begin{figure}[h]
    \centering
    \includegraphics[width=1\columnwidth]{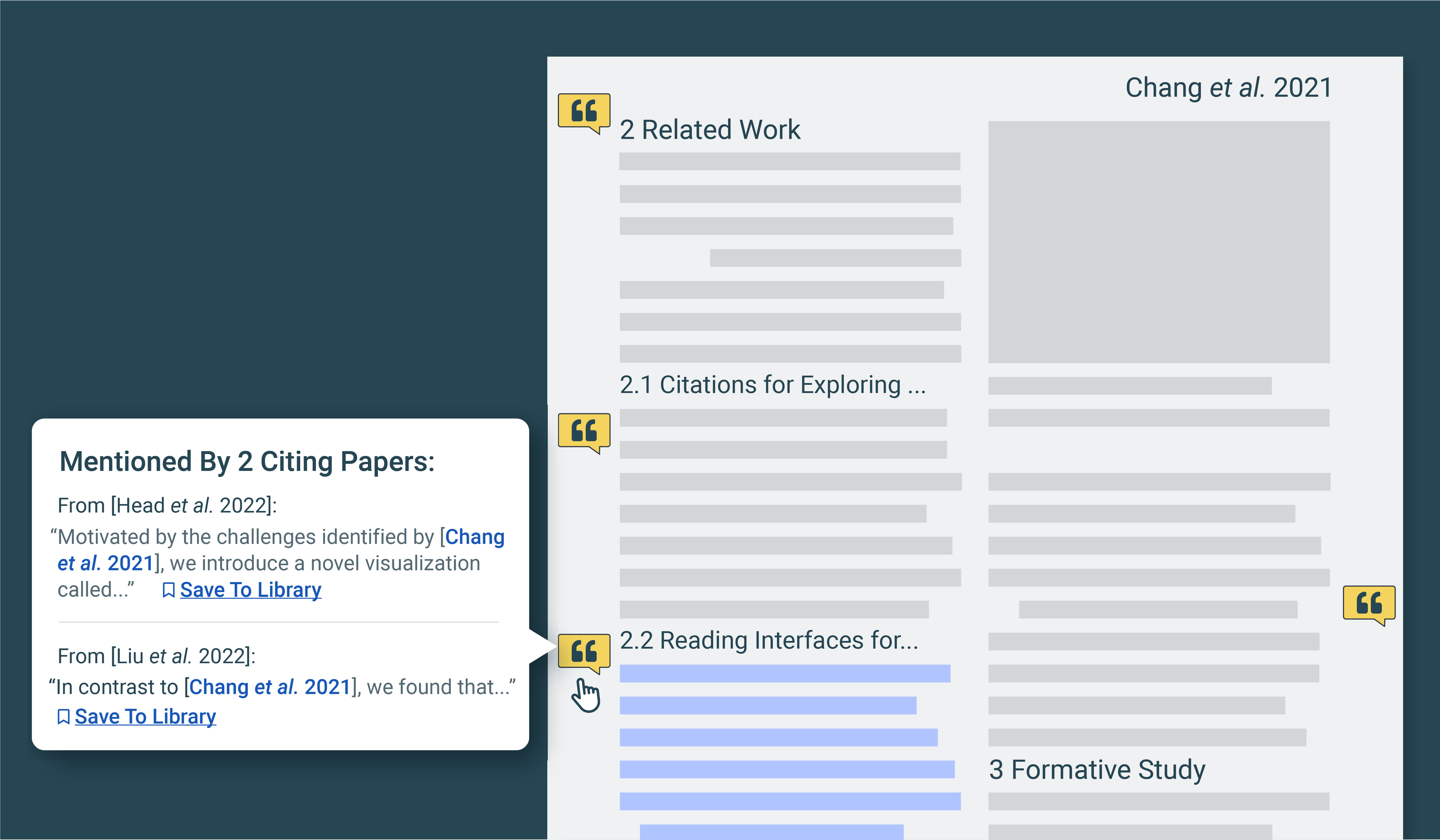}
    \caption{
   CiteRead \cite{Rachatasumrit2022CiteReadIL} finds subsequently published citing {\papers}, extracts the citation context, and localizes it to relevant parts of the current {\paper} as margin notes. This allows {\users} to become aware of important follow on work and explore them in-situ.
    }
    \label{fig:citation-citeread}
\end{figure}

\subsection{Exploring Future Work with CiteRead}
\label{sec:cite-read}

While augmenting inline citations helps readers to triage them, many relevant {\papers} are not cited in a {\paper} in the first place, for example, because they were published afterwards.
CiteRead is a novel {\system} that helps {\users} discover how follow-on work has built on or engaged with the {\paper} \cite{Rachatasumrit2022CiteReadIL}.
Much like social document annotation systems~\cite{zyto2012successful}, CiteRead annotates text in the paper with margin notes containing relevant commentary from citing papers~\cite{nakov2004citances}, thereby helping the reader to become aware of the citing paper and its connection.
In order to produce these annotations automatically, CiteRead first filters citing {\papers} for ones that are most relevant to the reader using a trained model atop a number of features representing citational discourse and textual similarity, i.e. from scientific paper embeddings~\cite{cohan2020specter}.
CiteRead then localizes citing papers to particular spans of text in the paper being read, and extracts relevant information from the citing paper.
Figure~\ref{fig:citation-citeread} shows a {\paper} annotated with this information from citing papers.
Localization is a technical challenge because while inline citations reference cited papers, they do not typically reference specific locations in the cited paper; CiteRead determines location by looking for overlapping spans of text (e.g., a number in common in the citing paper and the cited paper) or localizes to the relevant section when this overlap is unavailable. With CiteRead, a {\user} can directly examine follow-on work while keeping the citation contexts of both the current paper and the citing paper.
In a lab study, CiteRead helped readers better understand a {\paper} and its follow-on work compared to providing readers with a separate interface for faceted browsing of follow-on work. %

\section{Navigation and Efficient Reading}
\label{sec:guiding}

An important part of reading a paper is knowing what and where to read. Scholars often read papers non-linearly; they might return to a previously-read passage to recall some information, or jump forward to a different section of the paper (or to another paper) to satisfy an information need before jumping back. While jumping can help scholars orient their reading to sections of interest, it can also be a distraction by causing readers to constantly switch contexts. Non-linear navigation can be especially burdensome when the reader is interested in a particular \emph{type} of information (e.g., skimming a paper for the main results), but doesn't know precisely where to find it within the paper.
In this section we discuss two systems, Scim~\cite{Fok2023Scim} and Ocean~\cite{park-cscw22}, which demonstrate different approaches to helping readers navigate efficiently through a paper toward high-value, relevant information.

\subsection{Guided Reading with Scim}
\label{sec:scim}

\begin{figure}[h]
    \centering
    \includegraphics[width=1\columnwidth]{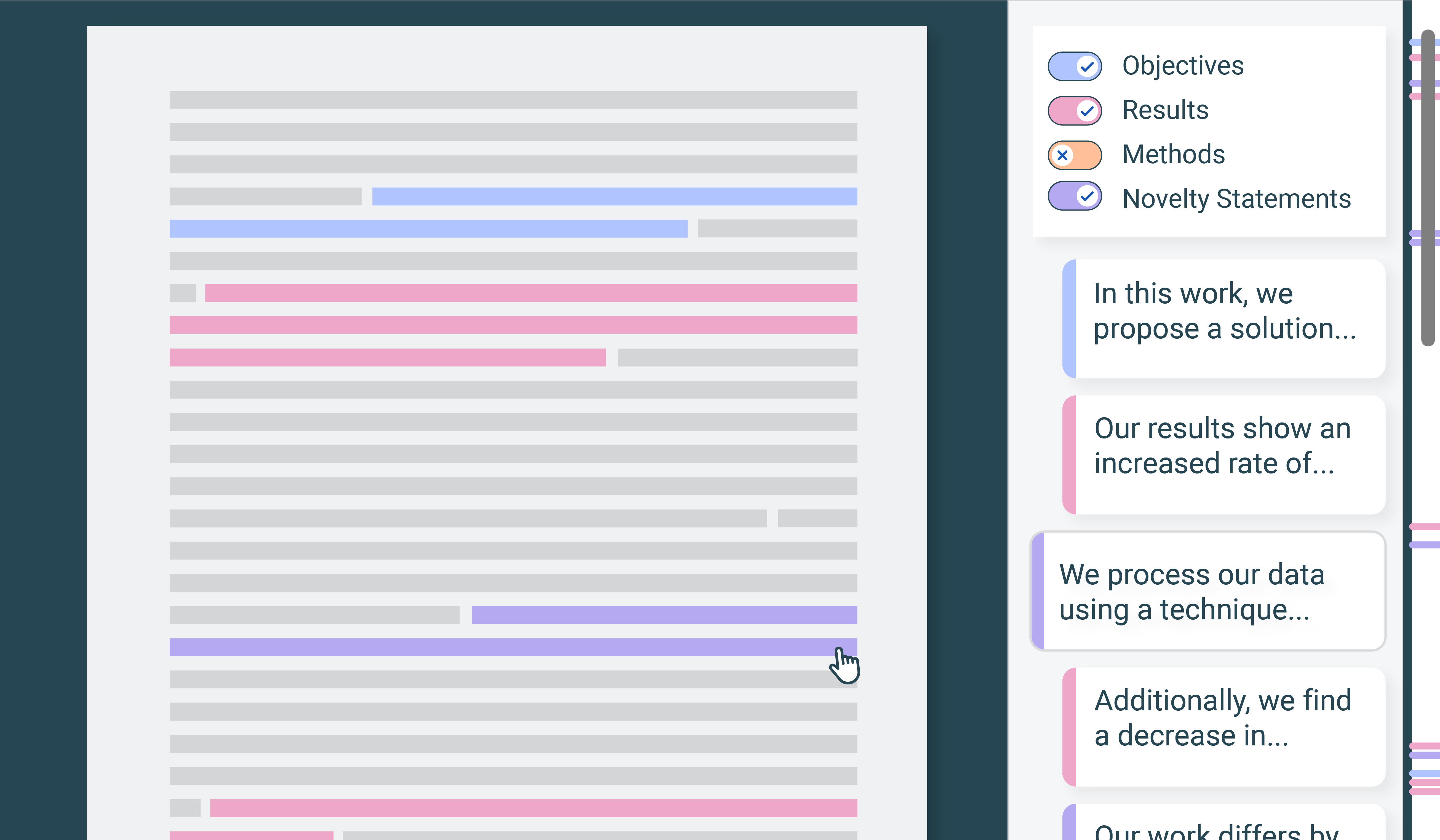}
    \caption{The Scim~\cite{Fok2023Scim} interface guides reader attention using color highlights corresponding to discourse facets. A sidebar allows users to toggle facets on/off. Clicking a color-coded snippet scrolls the reader to the relevant passage.
    }
    \label{fig:guided-reading}
\end{figure}

Scholarly reading can be considered a type of sensemaking represented as a continuous interplay between two processes: \textit{information foraging} in which readers identify relevant paper content, and \textit{comprehension} in which readers attempt to integrate the new information into their working model of the paper and with relevant prior knowledge~\cite{Pirolli_InformationForaging_1999, Russell_Sensemaking_1993}. Distinguishing between relevant and irrelevant content could help facilitate efficient reading. Paper abstracts offer one such separation, in essence an author-crafted determination of relevant content. However, static paper abstracts can leave readers to desire additional details that then require them to skim the paper itself.

Scim~\cite{Fok2023Scim} addresses this problem via an augmented reading interface designed to guide readers' attention using automatically-created in-situ faceted highlights (Figure \ref{fig:guided-reading}). Though prior work has explored highlighting as a visual cue for guiding reader attention~\cite{wecker_semantize_2014, chi_scenthighlights_2005, yang_hitext_2017}, the efficacy for reading of scholarly text is less well-understood. Scim investigated the following design goals for intelligent highlights in scholarly reading:  highlights should be (1) evenly-distributed throughout a paper, (2) have just the right density (too few highlights will present the guise of an inept tool, and too many will slow a reader down), and (3) highlight several key categories of information in the paper. Because readers often skim for common types of information, Scim uses a pretrained language model~\cite{wang-etal-2021-minilmv2} to classify salient sentences within papers into one of four information facets: research objectives, novel aspects of the research, methodology, and results, coupled with heuristics that ensure an even distribution of highlights. Usability studies of Scim have shown these highlights can reduce the time it takes readers to find specific information within a paper.
Readers found Scim particularly useful when skimming text-dense papers, or for papers that fell outside their area of expertise. Moreover, readers learned to use both Scim's inline highlights and a sidebar summary of highlights to augment their existing reading strategies.

\subsection{Low-Vision Navigation Support and Reader-Sourced Hyperlinks with Ocean}
\label{sec:ocean}
The task of navigating between sections and retrieving content can be particularly challenging for blind and low-vision readers due to limitations in auditory information access or small viewports under high magnification~\cite{Szpiro2016HowPW}.
Even when related content is linked, a small viewport can make navigation difficult and necessitate scrolling~\cite{park-cscw22}.
Most existing tools such as for auditory skimming~\cite{Khan2020DesigningAE} do not address such challenges associated with low-vision and magnification.

Ocean~\cite{park-cscw22} minimizes scrolling requirements for low-vision readers by providing bi-directional, viewport-preserving hyperlinks that enable navigating to and from associated content without disrupting the viewport.
Based on reported findings from interviews with low-vision readers, Ocean also allows for easily revisiting portions of the paper with tabbed reading.
Since papers do not always provide hyperlinks and automated link creation is imperfect, Ocean includes an authoring interface that allows readers to create and share paper links during reading. An exploratory field deployment study with mixed-ability groups of low-vision and sighted readers revealed that readers found value in creating and consuming these links, and that reader-created links can increase trust. %

\section{In-Situ Explanations for Better Comprehension}
\label{sec:in-situ-explanations}

Could an augmented reading application help readers understand a paper by reducing the  cognitive load associated with reading a paper? 
In this section, we discuss several ways in which interactive reading aids can help a reader understand a paper with less work through three systems: ScholarPhi~\cite{Head2021AugmentingSP}, PaperPlain~\cite{August2022PaperPM} and Papeo~\cite{papeo}. 
In particular, papers can be augmented with  definitions of terms and symbols,  provide plain-language summaries of paper passages, and connect readers with alternative forms of expression (for instance, video clips of research talks) that offer more approachable explanations of the paper's content.
\subsection{Defining Terms and Symbols with ScholarPhi}
\label{s:individual-terms}

\begin{figure}[b] 
     \centering
     \includegraphics[width=1\columnwidth]{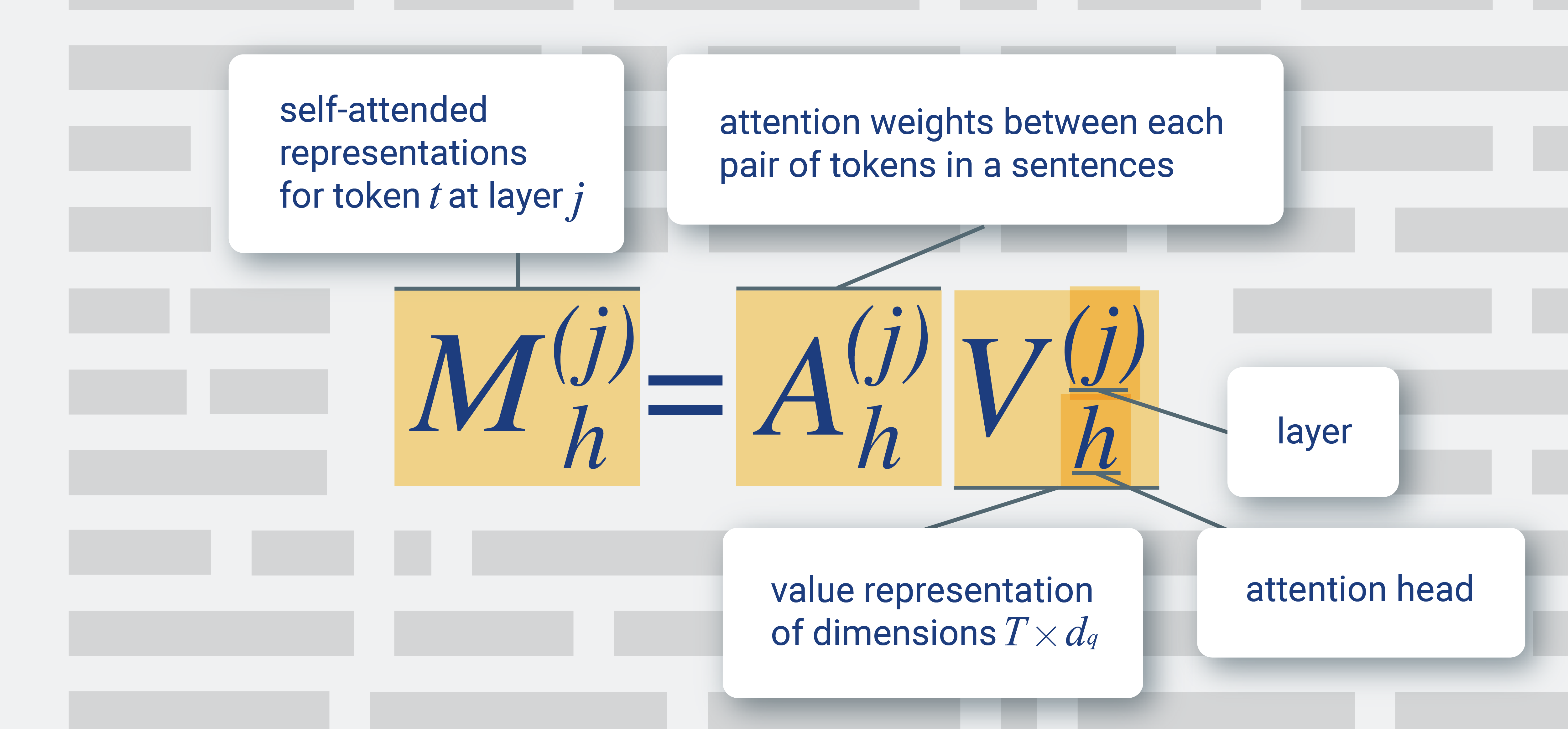}        \caption{ScholarPhi~\cite{Head2021AugmentingSP} shows definitions of terms and symbols in pop-up tooltips. When a reader selects a formula, all known definitions of symbols are shown simultaneously.
     To let readers select nested symbols (e.g., ``$h$'' in ``$V^{(j)}_{h}$''), ScholarPhi supports ``drill-down'' subsymbol selection.
     }
     \label{fig:scholarphi}
\end{figure}

Understanding a paper requires understanding its vocabulary. However, this is by no means an easy task---a typical paper may contain dozens of acronyms, symbols, and invented terms. And often, these terms appear without accompanying definitions~\cite{murthy2022accord}. 
How can we design interactive aids that present definitions of terms when and where readers most need them? ScholarPhi~\cite{Head2021AugmentingSP} takes as its basis the term gloss---an extension to a reading interface that shows a reader an explanation of a phrase when they click it. Glosses appeared in early research interfaces for reading hypertext~\cite{ref:zellweger1998fluid} and have since become part of widely-used reading interfaces including Wikipedia and Kindle.

That said, familiar gloss designs do not work well for scientific papers, where glosses run the risk of distracting readers, terms have multiple meanings, and phrases (specifically math symbols) are difficult to unambiguously select. 
The ScholarPhi design addresses these challenges. First, it aims to reduce distraction by showing definitions with high economy: glosses show multiple definitions and
 and in-context usages within a compact tooltip.
 Second, it provides position-sensitive definitions, revealing definitions that appears most recently prior to the selected usages of terms. 
 Terms and definitions are automatically identified using a pretrained language model~\cite{kang-etal-2020-document}.
 Finally, it provides easier access to definitions of mathematical symbols. Readers can access definitions of both a symbol and the subsymbols it is made of through a multi-click, ``drill-down'' selection mechanism. Furthermore, when a reader selects a formula, they can see definitions for all symbols at once, automatically placed adjacent to the symbols in the formula's margins (see Figure~\ref{fig:scholarphi}).
 
 In a usability study, the above interactions reduced the time it took readers to find answers to questions involving the understanding of terminology. All readers reported they would use the definition tooltips and formula diagrams often or always if available in their PDF reader tools.

\subsection{Simplifying Complex Passages with Paper Plain}
\label{sec:paper-plain}

\begin{figure}[t]
     \centering
     \includegraphics[width=1\columnwidth]{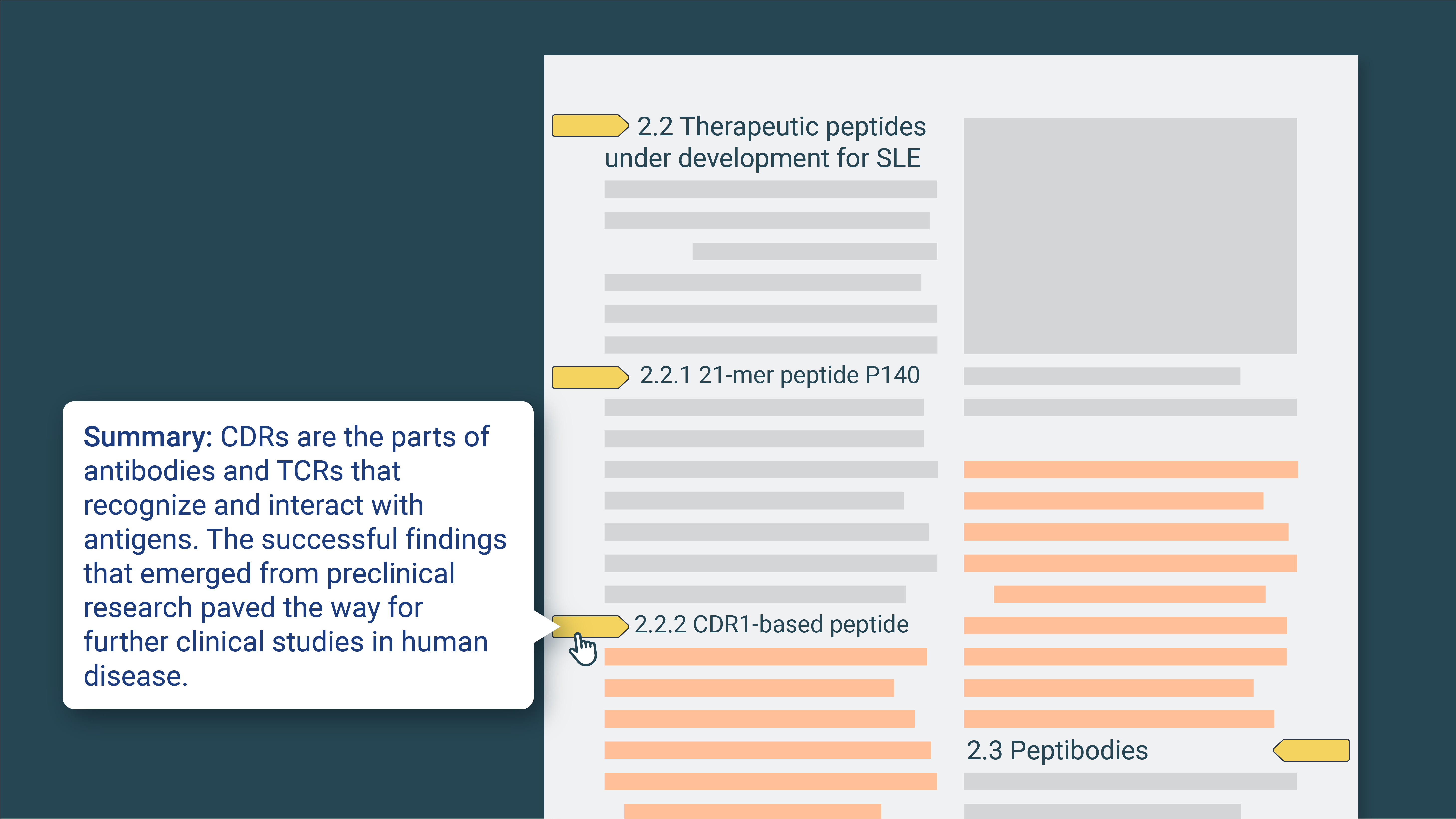}
     \caption{
     Paper Plain~\cite{August2022PaperPM} 
     provides in-situ plain language summaries of passages called ``gists'' to help readers who are overwhelmed by complex textual passages. Readers access gists by clicking a flag next to a section header. These gists are generated by large language models.  
     }
     \label{fig:paper-plain}
\end{figure}

Helping a reader understand individual terms and phrases only addresses part of the problem. Papers often contain passages so dense and complex that individual definitions are not enough to help someone read the passages, especially if they are a novice or non-expert in a field \cite{Britt2014ScientificLT}. 
Can we make complex texts more approachable by incorporating plain language summaries in the margins of the text?
With Paper Plain~\cite{August2022PaperPM}, when a reader encounters a section they find difficult to read, they can access a plain language summary of that section by clicking a button adjacent to the section header (see Figure~\ref{fig:paper-plain}). 
These summaries are generated by prompting a large language model with section text~\cite{gpt3-brown-2020}. 

Furthermore, Paper Plain helps guide readers using these summaries as an ``index'' into the text. A sidebar containing questions a reader may have about the text (e.g., \emph{What did the paper find?} or \emph{What were the limitations?}) provides links into answering passages identified using a question-answering system~\cite{Yoon2019PretrainedLM} alongside their associated plain language summaries.
These features were designed to help readers understand the ``gist'' of passages that contain unfamiliar vocabulary, providing support beyond that of individual term definitions.
Drawing inspiration from prior interactive reading affordances for term definitions~\cite{Jain2018ContentDE}, in-situ question answering~\cite{Zhao2020TalkTP, Chaudhri2013InquireBA}, and guiding reading~\cite{Dzara2019MedicalEJ}, Paper Plain seeks to bring these features together into a holistic system capable of supporting reading of a paper by a non-expert readership. In a usability study, readers made more frequent use of passage summaries than definition tooltips when both were available, suggesting the potential value of plain language summaries as allowing readers to bypass definitions of individual terms when acquiring a broad understanding of a paper.

\vspace{-2mm}

\subsection{Fusing Papers and Videos with Papeo}
\label{sec:papeo}

Sometimes, the best explanation of an idea is non-textual.
Videos can enhance understanding~\cite{Mayer1998ACT} while also requiring less mental load~\cite{Mayer1998ASE}, and various tools have been designed to facilitate searching and browsing for explanations in informational videos such as lectures~\cite{Kim2014DatadrivenIT, Liu2018ConceptScapeCC, Pavel2014VideoDA, Krosnick2015VideoDocC} and tutorials~\cite{Kim2014CrowdsourcingSI, Truong2021AutomaticGO, Khandwala2018CodemotionET}.
Similarly, for research papers, an algorithm might be better explained through an animation, a user interface might be better showcased through an screen recording, compared to the proses of a paper \cite{Hffler2007InstructionalAV}.
Instead of consuming the two formats independently, could interactive reading interfaces offer readers access to these alternative, more powerful descriptive forms as they read? 
For this, Papeo~\cite{papeo} was developed as a tool that supplements papers with more engaging, concise, dynamic presentations of information by linking excerpts of talk videos to corresponding paper passages.
 To grant authors more control over how their work is presented, we developed an AI-supported authoring interface for linking paper passages and videos efficiently: candidate passages are linked to excerpts of videos as suggestions using a pretrained language model~\cite{wang-2020-minilm-v1}, and an author interactively confirms or refines them. 
 
 Unlike text-skimming with Scim (\S\ref{sec:scim}) and Paper Plain (\S\ref{sec:paper-plain}), video-skimming in Papeo combines multiple modalities to explain complex information. For example, instead of reading a long text description of an interactive system, readers could see the system's behavior in a screen recording video with the author's commentary, and switch to corresponding passages to see implementation details or design motivations if desired.
Our early-stage evaluations of Papeo suggest that readers can use these interactions to fluidly transition between watching video and reading text, using video to quickly understand, and then selectively descending into the text when they desire a detailed understanding of the paper.

\section{Scaffolding Synthesis with Related Work Sections}
\label{sec:bootstrap}

Scientific breakthroughs often rely upon scholars synthesizing multiple published works into broad overviews  to identify gaps in the current literature \cite{portenoy2022bursting}. For this, scholars periodically compile survey articles to help other scholars gain a comprehensive overview of important research topics. For example, some fields have dedicated outlet for such articles (e.g., the \textit{Psychological Bulletin} \cite{bem1995writing}). However, survey articles require significant time and effort to synthesize, and can quickly become outdated with the exponential growth of scientific publication \cite{Bornmann2020GrowthRO}. 

Instead, scholars in fast-paced disciplines often rely on the related work section when they need to better understand the broader background when reading a paper. 
While related work sections also summarize multiple prior works, unlike comprehensive survey articles, they typically provide partial views of the larger research topic most relevant to a single paper. There is an opportunity to build better tooling for scholars to consume and synthesize multiple related work sections across many papers to gain richer and more comprehensive overviews of fast-paced domains. 
The Threddy \cite{Kang2022Threddy} and Related \cite{relatedly} projects explored this opportunity using two different approaches: clipping and organizing research threads mentioned across papers \cite{Kang2022Threddy}, and directly exploring and reading related work sections extracted across many papers \cite{relatedly}.

\begin{figure}[t] 
     \centering
     \includegraphics[width=1\columnwidth]{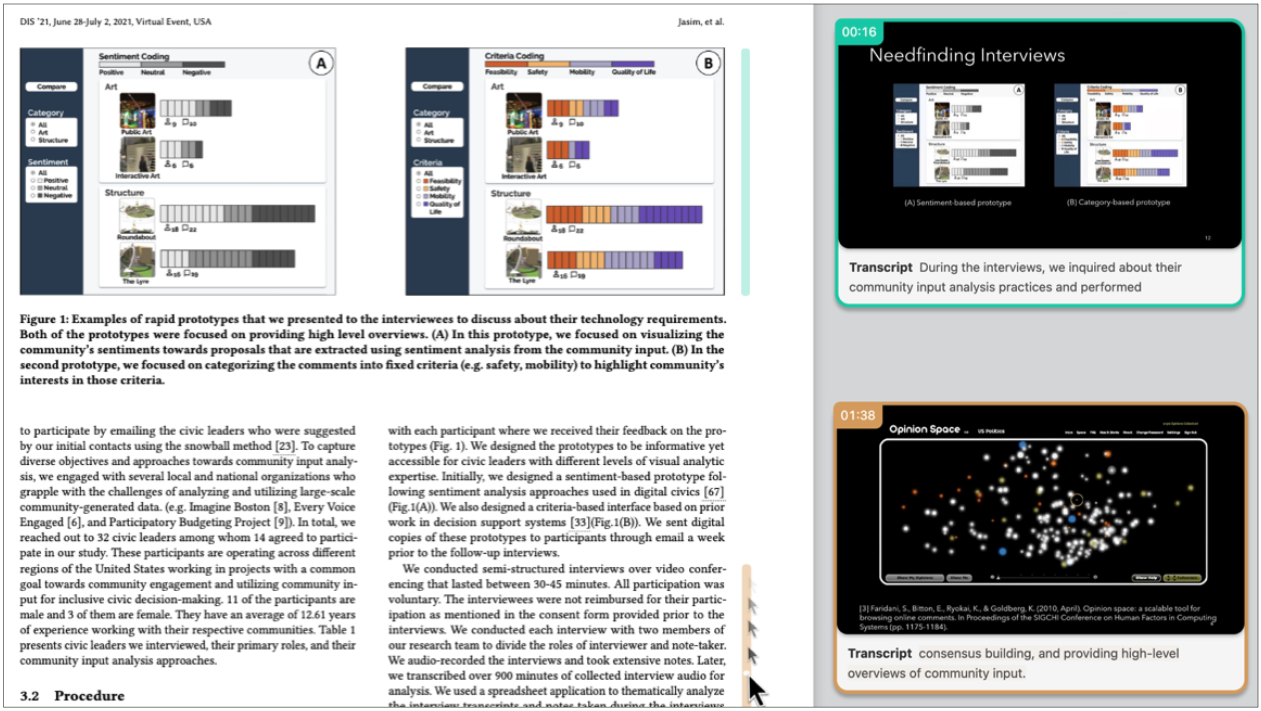}
     \caption{Papeo \cite{papeo} enables authors to map segments of talk videos to relevant passages in the paper, allowing readers to fluidly switch between the two formats. Color-coded bars show the mapping between the two formats, and allow readers to scrub through video segments for quick previews.
     }
     \label{fig:papeo}
\end{figure}

\subsection{Clipping and Synthesizing across Papers with Threddy}
\label{sec:threddy}

Clipping and note-taking is one common approach to supporting synthesis across multiple documents.
Prior work has pointed to the importance of tightly integrating clipping and synthesis support in the reading process, and how incurring significant context-switching costs can be detrimental to sensemaking~\cite{kittur_chi13_cost_benefit,Russell_Sensemaking_1993,Pirolli_InformationForaging_1999}.
Therefore, recent work has developed tools aimed at reducing the cognitive and interaction costs of clipping~
\cite{liu_wigglite,chang_uist16_uncertain_highlighting} and structuring~\cite{chang_mesh,kuznetsov_fuse,liu_crystalline,liu2019unakite,texSketch} to support everyday online researchers \cite{chang_mesh}, programmers \cite{liu2019unakite}, and students \cite{texSketch}. However, designing clipping and synthesis support tools for research papers is relatively under-explored and introduces exciting new research opportunities. For example, additional organizational structures for literature reviews (\eg threads of prior work instead of tables \cite{liu2019unakite,chang_mesh}), and research paper discovery (\eg based on inline citations in clipped text).

For this, Threddy \cite{Kang2022Threddy} is a thread-focused clipping tool integrated into scholars' paper reading process to support literature review and discovery. Using Threddy, readers can select and save sentences into a sidebar from the related work sections of a paper. The system maintains rich context for each clip, including its provenance and inline citations. This allows readers to navigate back to the clipped paper and cited papers afterward. In the sidebar, readers can further organize clips collected across papers into a hierarchy of threads to form their view of the research landscape. The content of the sidebar is preserved across papers that were read over time, and provides valuable context for subsequent reading based on the emerging threads of research the reader have curated.
Finally, readers can further expand their coverage by exploring paper recommendations for each thread, based on the referenced papers in the corresponding clips. 
A lab study showed that Threddy was able to lower the interaction costs of saving clips while maintaining context, allowed participants to curate research threads without breaking reading flows, and discover interesting new papers to further grow their understanding of the research fields.

\subsection{Reading and Exploring Related Work Sections across Papers with Relatedly}
\label{sec:relatedly}

In contrast to Threddy, which aims to improve readers' existing literature review process through enhanced in-situ clipping and synthesis \cite{Kang2022Threddy}, the Relatedly system introduced a novel workflow that allows readers to explore many related work sections across papers in an interactive search and reading interface to quickly gain a comprehensive overview of rich research topics \cite{relatedly}. While prior work have explored providing overview structure of multiple documents based on citations \cite{Ponsard2016PaperQuestAV,chau2011apolo}, semantic similarity \cite{hearst2006clustering,shahaf2012metro}, or human computation \cite{hahn2016knowledge,chang2016alloy,luther2015crowdlines}, they could still lead to complex structures that are hard to interpret \cite{hearst1999use} or require significant crowdsourcing efforts. Relatedly sidesteps these issues by \emph{reusing} existing related work paragraphs in published papers which already cite sets of related references with descriptions connecting them \cite{relatedly}. As an example, consider a scholar trying to better understand the space of \emph{online misinformation}. With \emph{online misinformation} as the query term, Relatedly shows the reader a list of paragraphs that describe and cite multiple relevant prior work. Using a pretrained language model for summarization~\cite{lewis-etal-2020-bart}, Relatedly generates short and descriptive titles for each paragraph, and uses a diversity-based ranking algorithm so that the reader can quickly see and explore paragraphs describing different research threads, such as \emph{Fact Checking Datasets}, \emph{Social Media and Misinformation}, and \emph{Fake News Detection Techniques}. 

One challenge here is that paragraphs of the same threads often cite overlapping prior work, making them hard to explore and read while keeping track of which papers were new versus already explored. For this, Relatedly provides reading and cross-referencing support by keeping track of paragraphs and references explored by the readers. This allows Relatedly to help readers prioritize their reading for both breadth and depth.
Specifically, Relatedly dynamically re-ranks paragraphs and highlights sentences to spotlight unexplored and dissimilar references for breadth, but also allow readers to explore clusters of paragraphs that cited similar references for depth.
A usability study comparing Relatedly to a strong document-centric baseline showed that Relatedly led to participants writing summaries that were rated significantly more coherent, insightful, and detailed after 20 minutes of literature review.

\section{Dynamic Documents for Improved Accessibility}
\label{sec:accessibility}

A range of disabilities cause people to read scientific documents using a wide variety of devices and reading tools.
For example, blind and low vision readers may use assistive reading technology such as screen readers, screen magnification, or text-to-speech to read documents~\cite{Szpiro2016HowPW}.
Furthermore, people without disabilities face situational impairments, such as the inability to view a screen while driving or may have a preference for consuming content on a small, mobile device. 

Many of these reading tools, such as screen readers, do not function properly on document formats designed for print such as PDF unless the document has been manually post-processed  to add information about reading order, content type, etc., which is rarely performed on scientific documents~\cite{bigham-uninteresting-tour,wang-2021-accessibility}. 
Further, certain content elements such as figures require the addition of alternative text in order to be read aloud at all (figure captions typically assume the reader can see the figure and do not provide the same semantic content as alt text).
High magnification reduces the viewport (the amount of visible content) and can dramatically increase the amount of scrolling and panning required, especially for multi-columnar formats that are commonly used by scientific documents. Visual scanning for information may be impacted or unavailable in these settings, making it more difficult to find and navigate between content in the document~\cite{park-cscw22}.

One way to render legacy PDF content more accessibly is to parse and convert it into a more flexible format, such as XML or HTML, which can then be formatted for mobile devices and augmented for reading by screen readers. The SciA11y system\footnote{A demo of a subsequent version is available at \url{https://papertohtml.org/}}
demonstrates this approach, automatically converting 12M academic PDFs to HTML~\cite{wang-2021-scia11y}; a user study with blind and low vision participants demonstrated strong user appreciation of the output, though some errors remain (e.g., failing in certain cases to distinguish footnotes from body text, difficulty parsing math equations)~\cite{wang-2021-accessibility}.
When available, alt text can be automatically categorized into semantic content types, enabling new reading experiences that allow skipping or prioritizing certain types~\cite{chintalapati-alt-text}.
Other approaches provide complementary benefits, such as interfaces tailored for low-vision readers (\S~\ref{sec:ocean}), as well as the range of reading support systems outlined above.

\section{Discussion and Future Work}

There are additional directions to explore to better support scholarly activities through the Semantic Reader Project. 

\paragraph{Towards a full-featured reading experience.} One question is how to integrate the different kinds of functionality across these projects into one coherent user interface, especially as we migrate research features into the production interface. 
Another question is how to develop support for the oftentimes social and collaborative nature of scholarly reading. Scholars frequently leverage their social networks and other social signals for paper discovery~\cite{kang_from_who_you_know}, work in groups to conduct literature review triage and synthesis, or engage in reading group discussions to aid comprehension. Existing augmentations within the Semantic Reader product could imbue social information, such as providing signals from one's co-author network (e.g., in CiteSee \S\ref{sec:citesee}) or aggregate navigation traces (e.g., in Scim \S\ref{sec:scim}). The publicly-available Semantic Reader tool could also scaffold the creation of novel crowd- or community-sourced content, such as author- or reader-provided explanations, commentary, or verification of paper content.
Finally is the question of how we can allow the scholarly community to step in where current AI systems fall short, such as by fixing improperly-extracted content or incorrect generated text which are especially problematic for interfaces  such as SciA11y (\S\ref{sec:accessibility}).

\paragraph{Advancing AI for scholarly documents.}
The Semantic Reader Project presents an opportunity for further AI research in scholarly document processing, especially when paired with human-centered research grounded in user-validated systems and scenarios. 
The bar for deploying AI models to support real-world reading is high; we often found during iterative design and usability studies that even slight errors in these models can have detrimental effects on the readers.
Until recently, interface design could require months of development of bespoke AI models which creates a barrier for quickly iterate different system designs.
Recent advancements in scaling large language models (LLMs) has altered this landscape by enabling researchers to experiment with a wide range of new NLP capabilities at relatively low cost~\cite{gpt3-brown-2020}.
This has the potential of significantly lowering the cost of human-centered AI design by incorporating user feedback in earlier stages of system development to create AI systems that work in symphony with the users beyond pure automation \cite{shneiderman2022human}.
For example, when developing Paper Plain (\S\ref{sec:paper-plain}), LLMs enabled us to quickly test different granularities and complexity-levels of plain language summaries with participants, eschewing the need for expensive changes to data requirements and model retraining.
In the near-term, we will revisit interface designs relying on bespoke AI models to evaluate whether LLMs can close the gap between research prototype and ready-for-production (e.g., more accurate definition identification for ScholarPhi \S\ref{s:individual-terms}).
Longer-term, we will explore whether LLMs can power new interactions (e.g., user-provided natural language queries while reading~\cite{dasigi-etal-2021-dataset,wadden-etal-2020-fact}).
While recent work has shown that these models can occasionally make critical errors or generate factually incorrect text when processing scientific text~\cite{otmakhova-etal-2022-patient},
we remain cautiously optimistic about developing ways to address their limitations \cite{dove2017ux,Lee2010GracefullyMB}.

\paragraph{Ethics of augmented papers}
Finally, all the new interfaces for reading that we propose pose a number of important ethical considerations that will require further research and discussion. One aspect that arises with any system for elevating certain papers or certain content over others is bias.
For instance, using signals such as citation counts faces the risk of a ``rich get richer'' bias, which can reflect other kinds of documented biases \cite{Beel2009GoogleSR,Maliniak2013TheGC,way2019productivity}. As a result, systems such as CiteSee (\S\ref{sec:citesee}) or Relatedly (\ref{sec:relatedly}) should carefully consider additional signals of relevance such as semantic similarity to surface newer and overlooked papers. 
Another tension that we have encountered is the potential discrepancy between author desires and reader desires for how a work is presented and how much control to provide authors. For instance, in our work on Papeo (\S\ref{sec:papeo}), we found that authors desired control over placement of their talk video snippets, even as they found automated mapping support to be helpful. In other cases, authors might not have the requisite expertise (e.g., they may not have a good sense of reader needs or what non-experts are confused by) or may have the wrong incentives.
Future work should consider author perspectives on these augmented experiences.
A related issue is around systems for more efficient reading or synthesis, which may  encourage readers to take shortcuts that lead to incorrect understanding, sloppy research, or even outright plagiarism. Instead of simply seeking to increase reading throughput uniformly, our systems should enable \textit{triage}, so that readers can dedicate time for thoughtful and careful reading when the content is important.
For instance, our systems could design pathways that, while they may be more efficient, do not obfuscate the full context (e.g., Scim \S\ref{sec:scim}), and that encourage good practices such as verification and provenance tracing.
A final consideration is around what is ethical reuse of a paper's contents to support reader experiences outside of that paper and its licensing implications. For instance, CiteRead (\S\ref{sec:cite-read}) extracts paper citances and places them in the cited paper, and Relatedly (\S\ref{sec:relatedly}) extracts related work sections from different papers for users to explore. 
Recent trends in \emph{open science and datasets} \cite{mckiernan2016open,mckiernan2000arxiv,ginsparg2011arxiv,lo-etal-2020-s2orc} point to a promising future where we could continue to explore different ways to \emph{remix and reuse} scholarly content across context so that future scientists can take fuller advantage of prior research.

\section{Conclusion}

This paper describes the Semantic Reader Project, which currently consists of ten research prototypes focusing on supporting scientists around Discovery \cite{Chang2022CiteSee,Rachatasumrit2022CiteReadIL}, Efficiency \cite{Fok2023Scim,park-cscw22}, Comprehension \cite{Head2021AugmentingSP,August2022PaperPM,papeo}, Synthesis \cite{relatedly, Kang2022Threddy}, and Accessibility \cite{wang-2021-scia11y,paper2html} when reading {\papers}.
Validating our approach of augmenting existing PDFs of {\papers}, we have seen tremendous adoption of the freely-available Semantic Reader product\footref{product} which has grown to 10k weekly users.\footnote{As of late February, 2023} 
While we focused on augmenting PDF documents to support common scholar reading practices, all of our {\systems} are built with web technologies---allowing these novel interactions to extend to future publication formats which can be rendered in web browsers. 
We plan to continue experimenting with novel AI-powered intelligent {\systems}, as well as migrating successful interactive features 
into the product. Finally, we offer a collection of freely-available resources to the larger research community, including datasets of open-access research papers \cite{lo-etal-2020-s2orc}, APIs for accessing the academic citation graph \cite{kinney2023semantic}, machine learning models for processing and understanding {\papers} \cite{cohan2020specter,Cachola2020TLDRES,shen-etal-2022-vila,kang-etal-2020-document},\footref{papermage} and open-source software for rendering and augmenting PDF documents for developing reading interfaces.\footref{library} We hope by providing these resources we can enable and encourage the broader research community to work on exciting novel intelligent {\systems} for {\papers} with us.

\begin{acks}
This project is supported in part by NSF Grant OIA-2033558, NSF Grant CNS-2213656. NSF RAPID Award 2040196, and ONR Grant N00014-21-1-2707.
\end{acks}

\bibliographystyle{ACM-Reference-Format}
\bibliography{anthology,reader,dan}


\begin{thebibliography}{109}


\ifx \showCODEN    \undefined \def \showCODEN     #1{\unskip}     \fi
\ifx \showDOI      \undefined \def \showDOI       #1{#1}\fi
\ifx \showISBNx    \undefined \def \showISBNx     #1{\unskip}     \fi
\ifx \showISBNxiii \undefined \def \showISBNxiii  #1{\unskip}     \fi
\ifx \showISSN     \undefined \def \showISSN      #1{\unskip}     \fi
\ifx \showLCCN     \undefined \def \showLCCN      #1{\unskip}     \fi
\ifx \shownote     \undefined \def \shownote      #1{#1}          \fi
\ifx \showarticletitle \undefined \def \showarticletitle #1{#1}   \fi
\ifx \showURL      \undefined \def \showURL       {\relax}        \fi
\providecommand\bibfield[2]{#2}
\providecommand\bibinfo[2]{#2}
\providecommand\natexlab[1]{#1}
\providecommand\showeprint[2][]{arXiv:#2}

\bibitem[August et~al\mbox{.}(2023)]%
        {August2022PaperPM}
\bibfield{author}{\bibinfo{person}{Tal August}, \bibinfo{person}{Lucy~Lu Wang},
  \bibinfo{person}{Jonathan Bragg}, \bibinfo{person}{Marti~A. Hearst},
  \bibinfo{person}{Andrew Head}, {and} \bibinfo{person}{Kyle Lo}.}
  \bibinfo{year}{2023}\natexlab{}.
\newblock \showarticletitle{Paper Plain: Making Medical Research Papers
  Approachable to Healthcare Consumers with Natural Language Processing}.
\newblock \bibinfo{journal}{\emph{ACM Transactions on Computer-Human
  Interaction}} (\bibinfo{year}{2023}).
\newblock
\newblock
\shownote{To appear}.


\bibitem[Bazerman(1985)]%
        {Bazerman1985-la}
\bibfield{author}{\bibinfo{person}{Charles Bazerman}.}
  \bibinfo{year}{1985}\natexlab{}.
\newblock \showarticletitle{Physicists reading physics: Schema-Laden Purposes
  and Purpose-Laden Schema}.
\newblock \bibinfo{journal}{\emph{Written Communication}} \bibinfo{volume}{2},
  \bibinfo{number}{1} (\bibinfo{date}{Jan.} \bibinfo{year}{1985}),
  \bibinfo{pages}{3--23}.
\newblock


\bibitem[Beel and Gipp(2009)]%
        {Beel2009GoogleSR}
\bibfield{author}{\bibinfo{person}{Joeran Beel} {and} \bibinfo{person}{Bela
  Gipp}.} \bibinfo{year}{2009}\natexlab{}.
\newblock \showarticletitle{Google Scholar's ranking algorithm: The impact of
  citation counts (An empirical study)}.
\newblock \bibinfo{journal}{\emph{2009 Third International Conference on
  Research Challenges in Information Science}} (\bibinfo{year}{2009}),
  \bibinfo{pages}{439--446}.
\newblock


\bibitem[Beltagy et~al\mbox{.}(2019)]%
        {Beltagy2019SciBERTAP}
\bibfield{author}{\bibinfo{person}{Iz Beltagy}, \bibinfo{person}{Kyle Lo},
  {and} \bibinfo{person}{Arman Cohan}.} \bibinfo{year}{2019}\natexlab{}.
\newblock \showarticletitle{SciBERT: A Pretrained Language Model for Scientific
  Text}. In \bibinfo{booktitle}{\emph{Conference on Empirical Methods in
  Natural Language Processing}}.
\newblock


\bibitem[Bem(1995)]%
        {bem1995writing}
\bibfield{author}{\bibinfo{person}{Daryl~J Bem}.}
  \bibinfo{year}{1995}\natexlab{}.
\newblock \showarticletitle{Writing a review article for Psychological
  Bulletin.}
\newblock \bibinfo{journal}{\emph{Psychological Bulletin}}
  \bibinfo{volume}{118}, \bibinfo{number}{2} (\bibinfo{year}{1995}),
  \bibinfo{pages}{172}.
\newblock


\bibitem[Bigham et~al\mbox{.}(2016)]%
        {bigham-uninteresting-tour}
\bibfield{author}{\bibinfo{person}{Jeffrey~P. Bigham}, \bibinfo{person}{Erin~L.
  Brady}, \bibinfo{person}{Cole Gleason}, \bibinfo{person}{Anhong Guo}, {and}
  \bibinfo{person}{David~A. Shamma}.} \bibinfo{year}{2016}\natexlab{}.
\newblock \showarticletitle{An Uninteresting Tour Through Why Our Research
  Papers Aren't Accessible}. In \bibinfo{booktitle}{\emph{Proceedings of the
  2016 CHI Conference Extended Abstracts on Human Factors in Computing
  Systems}} (San Jose, California, USA) \emph{(\bibinfo{series}{CHI EA '16})}.
  \bibinfo{publisher}{Association for Computing Machinery},
  \bibinfo{address}{New York, NY, USA}, \bibinfo{pages}{621–631}.
\newblock
\showISBNx{9781450340823}
\urldef\tempurl%
\url{https://doi.org/10.1145/2851581.2892588}
\showDOI{\tempurl}


\bibitem[Bornmann et~al\mbox{.}(2020)]%
        {Bornmann2020GrowthRO}
\bibfield{author}{\bibinfo{person}{Lutz Bornmann}, \bibinfo{person}{Ruediger
  Mutz}, {and} \bibinfo{person}{Robin Haunschild}.}
  \bibinfo{year}{2020}\natexlab{}.
\newblock \showarticletitle{Growth rates of modern science: a latent piecewise
  growth curve approach to model publication numbers from established and new
  literature databases}.
\newblock \bibinfo{journal}{\emph{Humanities and Social Sciences
  Communications}}  \bibinfo{volume}{8} (\bibinfo{year}{2020}),
  \bibinfo{pages}{1--15}.
\newblock


\bibitem[Brainard(2020)]%
        {brainard2020scientists}
\bibfield{author}{\bibinfo{person}{Jeffrey Brainard}.}
  \bibinfo{year}{2020}\natexlab{}.
\newblock \showarticletitle{Scientists are drowning in COVID-19 papers. Can new
  tools keep them afloat}.
\newblock \bibinfo{journal}{\emph{Science}} \bibinfo{volume}{13},
  \bibinfo{number}{10} (\bibinfo{year}{2020}), \bibinfo{pages}{1126}.
\newblock


\bibitem[Britt et~al\mbox{.}(2014)]%
        {Britt2014ScientificLT}
\bibfield{author}{\bibinfo{person}{Mary~Anne Britt}, \bibinfo{person}{Tobias
  Richter}, {and} \bibinfo{person}{Jean-François Rouet}.}
  \bibinfo{year}{2014}\natexlab{}.
\newblock \showarticletitle{Scientific Literacy: The Role of Goal-Directed
  Reading and Evaluation in Understanding Scientific Information}.
\newblock \bibinfo{journal}{\emph{Educational Psychologist}}
  \bibinfo{volume}{49} (\bibinfo{year}{2014}), \bibinfo{pages}{104 -- 122}.
\newblock


\bibitem[Brown et~al\mbox{.}(2020)]%
        {gpt3-brown-2020}
\bibfield{author}{\bibinfo{person}{Tom Brown}, \bibinfo{person}{Benjamin Mann},
  \bibinfo{person}{Nick Ryder}, \bibinfo{person}{Melanie Subbiah},
  \bibinfo{person}{Jared~D Kaplan}, \bibinfo{person}{Prafulla Dhariwal},
  \bibinfo{person}{Arvind Neelakantan}, \bibinfo{person}{Pranav Shyam},
  \bibinfo{person}{Girish Sastry}, \bibinfo{person}{Amanda Askell},
  \bibinfo{person}{Sandhini Agarwal}, \bibinfo{person}{Ariel Herbert-Voss},
  \bibinfo{person}{Gretchen Krueger}, \bibinfo{person}{Tom Henighan},
  \bibinfo{person}{Rewon Child}, \bibinfo{person}{Aditya Ramesh},
  \bibinfo{person}{Daniel Ziegler}, \bibinfo{person}{Jeffrey Wu},
  \bibinfo{person}{Clemens Winter}, \bibinfo{person}{Chris Hesse},
  \bibinfo{person}{Mark Chen}, \bibinfo{person}{Eric Sigler},
  \bibinfo{person}{Mateusz Litwin}, \bibinfo{person}{Scott Gray},
  \bibinfo{person}{Benjamin Chess}, \bibinfo{person}{Jack Clark},
  \bibinfo{person}{Christopher Berner}, \bibinfo{person}{Sam McCandlish},
  \bibinfo{person}{Alec Radford}, \bibinfo{person}{Ilya Sutskever}, {and}
  \bibinfo{person}{Dario Amodei}.} \bibinfo{year}{2020}\natexlab{}.
\newblock \showarticletitle{Language Models are Few-Shot Learners}. In
  \bibinfo{booktitle}{\emph{Advances in Neural Information Processing
  Systems}}, \bibfield{editor}{\bibinfo{person}{H.~Larochelle},
  \bibinfo{person}{M.~Ranzato}, \bibinfo{person}{R.~Hadsell},
  \bibinfo{person}{M.F. Balcan}, {and} \bibinfo{person}{H.~Lin}} (Eds.),
  Vol.~\bibinfo{volume}{33}. \bibinfo{publisher}{Curran Associates, Inc.},
  \bibinfo{pages}{1877--1901}.
\newblock
\urldef\tempurl%
\url{https://proceedings.neurips.cc/paper/2020/file/1457c0d6bfcb4967418bfb8ac142f64a-Paper.pdf}
\showURL{%
\tempurl}


\bibitem[Cachola et~al\mbox{.}(2020)]%
        {Cachola2020TLDRES}
\bibfield{author}{\bibinfo{person}{Isabel Cachola}, \bibinfo{person}{Kyle Lo},
  \bibinfo{person}{Arman Cohan}, {and} \bibinfo{person}{Daniel~S. Weld}.}
  \bibinfo{year}{2020}\natexlab{}.
\newblock \showarticletitle{TLDR: Extreme Summarization of Scientific
  Documents}. In \bibinfo{booktitle}{\emph{Findings of EMNLP}}.
\newblock


\bibitem[Chang et~al\mbox{.}(2016a)]%
        {chang_uist16_uncertain_highlighting}
\bibfield{author}{\bibinfo{person}{Joseph~Chee Chang}, \bibinfo{person}{Nathan
  Hahn}, {and} \bibinfo{person}{Aniket Kittur}.}
  \bibinfo{year}{2016}\natexlab{a}.
\newblock \showarticletitle{Supporting Mobile Sensemaking Through Intentionally
  Uncertain Highlighting}. In \bibinfo{booktitle}{\emph{Proceedings of the 29th
  Annual Symposium on User Interface Software and Technology}} (Tokyo, Japan)
  \emph{(\bibinfo{series}{UIST '16})}. \bibinfo{publisher}{Association for
  Computing Machinery}, \bibinfo{address}{New York, NY, USA},
  \bibinfo{pages}{61–68}.
\newblock
\showISBNx{9781450341899}
\urldef\tempurl%
\url{https://doi.org/10.1145/2984511.2984538}
\showDOI{\tempurl}


\bibitem[Chang et~al\mbox{.}(2020)]%
        {chang_mesh}
\bibfield{author}{\bibinfo{person}{Joseph~Chee Chang}, \bibinfo{person}{Nathan
  Hahn}, {and} \bibinfo{person}{Aniket Kittur}.}
  \bibinfo{year}{2020}\natexlab{}.
\newblock \showarticletitle{Mesh: Scaffolding Comparison Tables for Online
  Decision Making}. In \bibinfo{booktitle}{\emph{Proceedings of the 33rd Annual
  ACM Symposium on User Interface Software and Technology}} (Virtual Event,
  USA) \emph{(\bibinfo{series}{UIST '20})}. \bibinfo{publisher}{Association for
  Computing Machinery}, \bibinfo{address}{New York, NY, USA},
  \bibinfo{pages}{391–405}.
\newblock
\showISBNx{9781450375146}
\urldef\tempurl%
\url{https://doi.org/10.1145/3379337.3415865}
\showDOI{\tempurl}


\bibitem[Chang et~al\mbox{.}(2016b)]%
        {chang2016alloy}
\bibfield{author}{\bibinfo{person}{Joseph~Chee Chang}, \bibinfo{person}{Aniket
  Kittur}, {and} \bibinfo{person}{Nathan Hahn}.}
  \bibinfo{year}{2016}\natexlab{b}.
\newblock \showarticletitle{Alloy: Clustering with crowds and computation}. In
  \bibinfo{booktitle}{\emph{Proceedings of the 2016 CHI Conference on Human
  Factors in Computing Systems}}. \bibinfo{pages}{3180--3191}.
\newblock


\bibitem[Chang et~al\mbox{.}(2023)]%
        {Chang2022CiteSee}
\bibfield{author}{\bibinfo{person}{Joseph~Chee Chang}, \bibinfo{person}{Amy~X
  Zhang}, \bibinfo{person}{Jonathan Bragg}, \bibinfo{person}{Andrew Head},
  \bibinfo{person}{Kyle Lo}, \bibinfo{person}{Doug Downey}, {and}
  \bibinfo{person}{Daniel~S Weld}.} \bibinfo{year}{2023}\natexlab{}.
\newblock \showarticletitle{CiteSee: Augmenting Citations in Scientific Papers
  with Persistent and Personalized Historical Context}.
\newblock \bibinfo{journal}{\emph{ArXiv}}  \bibinfo{volume}{abs/2022.99999}
  (\bibinfo{year}{2023}).
\newblock


\bibitem[Chau et~al\mbox{.}(2011)]%
        {chau2011apolo}
\bibfield{author}{\bibinfo{person}{Duen~Horng Chau}, \bibinfo{person}{Aniket
  Kittur}, \bibinfo{person}{Jason~I Hong}, {and} \bibinfo{person}{Christos
  Faloutsos}.} \bibinfo{year}{2011}\natexlab{}.
\newblock \showarticletitle{Apolo: making sense of large network data by
  combining rich user interaction and machine learning}. In
  \bibinfo{booktitle}{\emph{Proceedings of the SIGCHI conference on human
  factors in computing systems}}. \bibinfo{pages}{167--176}.
\newblock


\bibitem[Chaudhri et~al\mbox{.}(2013)]%
        {Chaudhri2013InquireBA}
\bibfield{author}{\bibinfo{person}{Vinay~K. Chaudhri},
  \bibinfo{person}{Britte~Haugan Cheng}, \bibinfo{person}{Adam Overholtzer},
  \bibinfo{person}{Jeremy Roschelle}, \bibinfo{person}{Aaron Spaulding},
  \bibinfo{person}{Peter~E. Clark}, \bibinfo{person}{Mark~T. Greaves}, {and}
  \bibinfo{person}{David Gunning}.} \bibinfo{year}{2013}\natexlab{}.
\newblock \showarticletitle{Inquire Biology: A Textbook that Answers
  Questions}.
\newblock \bibinfo{journal}{\emph{AI Mag.}}  \bibinfo{volume}{34}
  (\bibinfo{year}{2013}), \bibinfo{pages}{55--72}.
\newblock


\bibitem[Chi et~al\mbox{.}(2005)]%
        {chi_scenthighlights_2005}
\bibfield{author}{\bibinfo{person}{Ed~H. Chi}, \bibinfo{person}{Lichan Hong},
  \bibinfo{person}{Michelle Gumbrecht}, {and} \bibinfo{person}{Stuart~K.
  Card}.} \bibinfo{year}{2005}\natexlab{}.
\newblock \showarticletitle{ScentHighlights: highlighting conceptually-related
  sentences during reading}. In \bibinfo{booktitle}{\emph{Proceedings of the
  10th International Conference on Intelligent User Interfaces}}.
  \bibinfo{publisher}{Association for Computing Machinery},
  \bibinfo{address}{San Diego, CA, USA}, \bibinfo{pages}{272}.
\newblock


\bibitem[Chintalapati et~al\mbox{.}(2022)]%
        {chintalapati-alt-text}
\bibfield{author}{\bibinfo{person}{Sanjana~Shivani Chintalapati},
  \bibinfo{person}{Jonathan Bragg}, {and} \bibinfo{person}{Lucy~Lu Wang}.}
  \bibinfo{year}{2022}\natexlab{}.
\newblock \showarticletitle{A Dataset of Alt Texts from HCI Publications:
  Analyses and Uses Towards Producing More Descriptive Alt Texts of Data
  Visualizations in Scientific Papers}. In
  \bibinfo{booktitle}{\emph{Proceedings of the 24th International ACM SIGACCESS
  Conference on Computers and Accessibility}} (Athens, Greece)
  \emph{(\bibinfo{series}{ASSETS '22})}. \bibinfo{publisher}{Association for
  Computing Machinery}, \bibinfo{address}{New York, NY, USA}, Article
  \bibinfo{articleno}{30}, \bibinfo{numpages}{12}~pages.
\newblock
\showISBNx{9781450392587}
\urldef\tempurl%
\url{https://doi.org/10.1145/3517428.3544796}
\showDOI{\tempurl}


\bibitem[Cohan et~al\mbox{.}(2020)]%
        {cohan2020specter}
\bibfield{author}{\bibinfo{person}{Arman Cohan}, \bibinfo{person}{Sergey
  Feldman}, \bibinfo{person}{Iz Beltagy}, \bibinfo{person}{Doug Downey}, {and}
  \bibinfo{person}{Daniel~S Weld}.} \bibinfo{year}{2020}\natexlab{}.
\newblock \showarticletitle{Specter: Document-level representation learning
  using citation-informed transformers}.
\newblock \bibinfo{journal}{\emph{arXiv preprint arXiv:2004.07180}}
  (\bibinfo{year}{2020}).
\newblock


\bibitem[Conlen and Heer(2022)]%
        {Conlen2022FidyllAC}
\bibfield{author}{\bibinfo{person}{Matthew Conlen} {and}
  \bibinfo{person}{Jeffrey Heer}.} \bibinfo{year}{2022}\natexlab{}.
\newblock \showarticletitle{Fidyll: A Compiler for Cross-Format Data Stories \&
  Explorable Explanations}.
\newblock \bibinfo{journal}{\emph{ArXiv}}  \bibinfo{volume}{abs/2205.09858}
  (\bibinfo{year}{2022}).
\newblock


\bibitem[Conlen et~al\mbox{.}(2021)]%
        {conlen2021idyll}
\bibfield{author}{\bibinfo{person}{Matthew Conlen}, \bibinfo{person}{Megan Vo},
  \bibinfo{person}{Alan Tan}, {and} \bibinfo{person}{Jeffrey Heer}.}
  \bibinfo{year}{2021}\natexlab{}.
\newblock \showarticletitle{Idyll studio: A structured editor for authoring
  interactive \& data-driven articles}. In \bibinfo{booktitle}{\emph{The 34th
  Annual ACM Symposium on User Interface Software and Technology}}.
  \bibinfo{pages}{1--12}.
\newblock


\bibitem[Dasigi et~al\mbox{.}(2021)]%
        {dasigi-etal-2021-dataset}
\bibfield{author}{\bibinfo{person}{Pradeep Dasigi}, \bibinfo{person}{Kyle Lo},
  \bibinfo{person}{Iz Beltagy}, \bibinfo{person}{Arman Cohan},
  \bibinfo{person}{Noah~A. Smith}, {and} \bibinfo{person}{Matt Gardner}.}
  \bibinfo{year}{2021}\natexlab{}.
\newblock \showarticletitle{A Dataset of Information-Seeking Questions and
  Answers Anchored in Research Papers}. In
  \bibinfo{booktitle}{\emph{Proceedings of the 2021 Conference of the North
  American Chapter of the Association for Computational Linguistics: Human
  Language Technologies}}. \bibinfo{publisher}{Association for Computational
  Linguistics}, \bibinfo{address}{Online}, \bibinfo{pages}{4599--4610}.
\newblock
\urldef\tempurl%
\url{https://doi.org/10.18653/v1/2021.naacl-main.365}
\showDOI{\tempurl}


\bibitem[Denney and Tewksbury(2013)]%
        {Denney2013HowTW}
\bibfield{author}{\bibinfo{person}{Andrew~S. Denney} {and}
  \bibinfo{person}{Richard~Allan Tewksbury}.} \bibinfo{year}{2013}\natexlab{}.
\newblock \showarticletitle{How to Write a Literature Review}.
\newblock \bibinfo{journal}{\emph{Journal of Criminal Justice Education}}
  \bibinfo{volume}{24} (\bibinfo{year}{2013}), \bibinfo{pages}{218 -- 234}.
\newblock


\bibitem[Dove et~al\mbox{.}(2017)]%
        {dove2017ux}
\bibfield{author}{\bibinfo{person}{Graham Dove}, \bibinfo{person}{Kim Halskov},
  \bibinfo{person}{Jodi Forlizzi}, {and} \bibinfo{person}{John Zimmerman}.}
  \bibinfo{year}{2017}\natexlab{}.
\newblock \showarticletitle{UX design innovation: Challenges for working with
  machine learning as a design material}. In
  \bibinfo{booktitle}{\emph{Proceedings of the 2017 chi conference on human
  factors in computing systems}}. \bibinfo{pages}{278--288}.
\newblock


\bibitem[Dzara and Frey-Vogel(2019)]%
        {Dzara2019MedicalEJ}
\bibfield{author}{\bibinfo{person}{Kristina Dzara} {and}
  \bibinfo{person}{Ariel~S Frey-Vogel}.} \bibinfo{year}{2019}\natexlab{}.
\newblock \showarticletitle{Medical Education Journal Club for the Millennial
  Resident: An Interactive, No-Prep Approach.}
\newblock \bibinfo{journal}{\emph{Academic pediatrics}} (\bibinfo{year}{2019}).
\newblock


\bibitem[{Editorial Team}(2021)]%
        {Distill_Editorial_Team2021-ix}
\bibfield{author}{\bibinfo{person}{{Editorial Team}}.}
  \bibinfo{year}{2021}\natexlab{}.
\newblock \showarticletitle{Distill Hiatus}.
\newblock \bibinfo{journal}{\emph{Distill}} \bibinfo{volume}{6},
  \bibinfo{number}{7} (\bibinfo{date}{July} \bibinfo{year}{2021}).
\newblock
\urldef\tempurl%
\url{https://doi.org/10.23915/distill.00031}
\showDOI{\tempurl}


\bibitem[Fok et~al\mbox{.}(2023)]%
        {Fok2023Scim}
\bibfield{author}{\bibinfo{person}{Raymond Fok}, \bibinfo{person}{Hita
  Kambhamettu}, \bibinfo{person}{Luca Soldaini}, \bibinfo{person}{Jonathan
  Bragg}, \bibinfo{person}{Kyle Lo}, \bibinfo{person}{Andrew Head},
  \bibinfo{person}{Marti~A. Hearst}, {and} \bibinfo{person}{Daniel~S. Weld}.}
  \bibinfo{year}{2023}\natexlab{}.
\newblock \showarticletitle{Scim: Intelligent Skimming Support for Scientific
  Papers}. In \bibinfo{booktitle}{\emph{28th Annual Conference on Intelligent
  User Interfaces (IUI '23)}}.
\newblock


\bibitem[Ginsparg(2011)]%
        {ginsparg2011arxiv}
\bibfield{author}{\bibinfo{person}{Paul Ginsparg}.}
  \bibinfo{year}{2011}\natexlab{}.
\newblock \showarticletitle{ArXiv at 20}.
\newblock \bibinfo{journal}{\emph{Nature}} \bibinfo{volume}{476},
  \bibinfo{number}{7359} (\bibinfo{year}{2011}), \bibinfo{pages}{145--147}.
\newblock


\bibitem[Gori and Pucci(2006)]%
        {Gori2006ResearchPR}
\bibfield{author}{\bibinfo{person}{Marco Gori} {and} \bibinfo{person}{Augusto
  Pucci}.} \bibinfo{year}{2006}\natexlab{}.
\newblock \showarticletitle{Research Paper Recommender Systems: A Random-Walk
  Based Approach}.
\newblock \bibinfo{journal}{\emph{2006 IEEE/WIC/ACM International Conference on
  Web Intelligence (WI 2006 Main Conference Proceedings)(WI'06)}}
  (\bibinfo{year}{2006}), \bibinfo{pages}{778--781}.
\newblock


\bibitem[Hahn et~al\mbox{.}(2016)]%
        {hahn2016knowledge}
\bibfield{author}{\bibinfo{person}{Nathan Hahn}, \bibinfo{person}{Joseph
  Chang}, \bibinfo{person}{Ji~Eun Kim}, {and} \bibinfo{person}{Aniket Kittur}.}
  \bibinfo{year}{2016}\natexlab{}.
\newblock \showarticletitle{The Knowledge Accelerator: Big picture thinking in
  small pieces}. In \bibinfo{booktitle}{\emph{Proceedings of the 2016 CHI
  Conference on Human Factors in Computing Systems}}.
  \bibinfo{pages}{2258--2270}.
\newblock


\bibitem[He et~al\mbox{.}(2019)]%
        {He2019PaperPolesFA}
\bibfield{author}{\bibinfo{person}{Jiangen He}, \bibinfo{person}{Q. Ping},
  \bibinfo{person}{Wen Lou}, {and} \bibinfo{person}{Chaomei Chen}.}
  \bibinfo{year}{2019}\natexlab{}.
\newblock \showarticletitle{PaperPoles: Facilitating adaptive visual
  exploration of scientific publications by citation links}.
\newblock \bibinfo{journal}{\emph{Journal of the Association for Information
  Science and Technology}}  \bibinfo{volume}{70} (\bibinfo{year}{2019}).
\newblock


\bibitem[Head et~al\mbox{.}(2021)]%
        {Head2021AugmentingSP}
\bibfield{author}{\bibinfo{person}{Andrew Head}, \bibinfo{person}{Kyle Lo},
  \bibinfo{person}{Dongyeop Kang}, \bibinfo{person}{Raymond Fok},
  \bibinfo{person}{Sam Skjonsberg}, \bibinfo{person}{Daniel~S. Weld}, {and}
  \bibinfo{person}{Marti~A. Hearst}.} \bibinfo{year}{2021}\natexlab{}.
\newblock \showarticletitle{Augmenting Scientific Papers with Just-in-Time,
  Position-Sensitive Definitions of Terms and Symbols}.
\newblock \bibinfo{journal}{\emph{Proceedings of the 2021 CHI Conference on
  Human Factors in Computing Systems}} (\bibinfo{year}{2021}).
\newblock


\bibitem[Head et~al\mbox{.}(2022)]%
        {head2022math}
\bibfield{author}{\bibinfo{person}{Andrew Head}, \bibinfo{person}{Amber Xie},
  {and} \bibinfo{person}{Marti~A Hearst}.} \bibinfo{year}{2022}\natexlab{}.
\newblock \showarticletitle{Math Augmentation: How Authors Enhance the
  Readability of Formulas using Novel Visual Design Practices}. In
  \bibinfo{booktitle}{\emph{CHI Conference on Human Factors in Computing
  Systems}}. \bibinfo{pages}{1--18}.
\newblock


\bibitem[Hearst(1999)]%
        {hearst1999use}
\bibfield{author}{\bibinfo{person}{Marti~A Hearst}.}
  \bibinfo{year}{1999}\natexlab{}.
\newblock \showarticletitle{The use of categories and clusters for organizing
  retrieval results}.
\newblock \bibinfo{journal}{\emph{Natural language information retrieval}}
  (\bibinfo{year}{1999}), \bibinfo{pages}{333--374}.
\newblock


\bibitem[Hearst(2006)]%
        {hearst2006clustering}
\bibfield{author}{\bibinfo{person}{Marti~A Hearst}.}
  \bibinfo{year}{2006}\natexlab{}.
\newblock \showarticletitle{Clustering versus faceted categories for
  information exploration}.
\newblock \bibinfo{journal}{\emph{Commun. ACM}} \bibinfo{volume}{49},
  \bibinfo{number}{4} (\bibinfo{year}{2006}), \bibinfo{pages}{59--61}.
\newblock


\bibitem[H{\"o}ffler and Leutner(2007)]%
        {Hffler2007InstructionalAV}
\bibfield{author}{\bibinfo{person}{Tim~Niclas H{\"o}ffler} {and}
  \bibinfo{person}{Detlev Leutner}.} \bibinfo{year}{2007}\natexlab{}.
\newblock \showarticletitle{Instructional animation versus static pictures: A
  meta-analysis}.
\newblock \bibinfo{journal}{\emph{Learning and Instruction}}
  \bibinfo{volume}{17} (\bibinfo{year}{2007}), \bibinfo{pages}{722--738}.
\newblock


\bibitem[Hohman et~al\mbox{.}(2020)]%
        {Hohman2020CommunicatingWI}
\bibfield{author}{\bibinfo{person}{Fred Hohman}, \bibinfo{person}{Matthew
  Conlen}, \bibinfo{person}{Jeffrey Heer}, {and} \bibinfo{person}{Duen~Horng
  Chau}.} \bibinfo{year}{2020}\natexlab{}.
\newblock \showarticletitle{Communicating with Interactive Articles}.
\newblock \bibinfo{journal}{\emph{Distill}}.
\newblock
\urldef\tempurl%
\url{https://doi.org/10.23915/distill.00028}
\showDOI{\tempurl}


\bibitem[Huang et~al\mbox{.}(2022)]%
        {Huang2022LayoutLMv3PF}
\bibfield{author}{\bibinfo{person}{Yupan Huang}, \bibinfo{person}{Tengchao Lv},
  \bibinfo{person}{Lei Cui}, \bibinfo{person}{Yutong Lu}, {and}
  \bibinfo{person}{Furu Wei}.} \bibinfo{year}{2022}\natexlab{}.
\newblock \showarticletitle{LayoutLMv3: Pre-training for Document AI with
  Unified Text and Image Masking}.
\newblock \bibinfo{journal}{\emph{Proceedings of the 30th ACM International
  Conference on Multimedia}} (\bibinfo{year}{2022}).
\newblock


\bibitem[Huang et~al\mbox{.}(2002)]%
        {Huang2002AGR}
\bibfield{author}{\bibinfo{person}{Zan Huang}, \bibinfo{person}{Wingyan Chung},
  \bibinfo{person}{Thian-Huat Ong}, {and} \bibinfo{person}{Hsinchun Chen}.}
  \bibinfo{year}{2002}\natexlab{}.
\newblock \showarticletitle{A graph-based recommender system for digital
  library}. In \bibinfo{booktitle}{\emph{JCDL '02}}.
\newblock


\bibitem[Jain et~al\mbox{.}(2018)]%
        {Jain2018ContentDE}
\bibfield{author}{\bibinfo{person}{Abhinav Jain}, \bibinfo{person}{Nitin
  Gupta}, \bibinfo{person}{Shashank Mujumdar}, \bibinfo{person}{Sameep Mehta},
  {and} \bibinfo{person}{Rishi Madhok}.} \bibinfo{year}{2018}\natexlab{}.
\newblock \showarticletitle{Content Driven Enrichment of Formal Text using
  Concept Definitions and Applications}.
\newblock \bibinfo{journal}{\emph{Proceedings of the 29th on Hypertext and
  Social Media}} (\bibinfo{year}{2018}).
\newblock


\bibitem[Kang et~al\mbox{.}(2020)]%
        {kang-etal-2020-document}
\bibfield{author}{\bibinfo{person}{Dongyeop Kang}, \bibinfo{person}{Andrew
  Head}, \bibinfo{person}{Risham Sidhu}, \bibinfo{person}{Kyle Lo},
  \bibinfo{person}{Daniel Weld}, {and} \bibinfo{person}{Marti~A. Hearst}.}
  \bibinfo{year}{2020}\natexlab{}.
\newblock \showarticletitle{Document-Level Definition Detection in Scholarly
  Documents: Existing Models, Error Analyses, and Future Directions}. In
  \bibinfo{booktitle}{\emph{Proceedings of the First Workshop on Scholarly
  Document Processing}}. \bibinfo{publisher}{Association for Computational
  Linguistics}, \bibinfo{address}{Online}, \bibinfo{pages}{196--206}.
\newblock
\urldef\tempurl%
\url{https://doi.org/10.18653/v1/2020.sdp-1.22}
\showDOI{\tempurl}


\bibitem[Kang et~al\mbox{.}(2022a)]%
        {Kang2022Threddy}
\bibfield{author}{\bibinfo{person}{Hyeonsu~B Kang},
  \bibinfo{person}{Joseph~Chee Chang}, \bibinfo{person}{Yongsung Kim}, {and}
  \bibinfo{person}{Aniket Kittur}.} \bibinfo{year}{2022}\natexlab{a}.
\newblock \showarticletitle{Threddy: An Interactive System for Personalized
  Thread-based Exploration and Organization of Scientific Literature}.
\newblock \bibinfo{journal}{\emph{arXiv preprint arXiv:2208.03455}}
  (\bibinfo{year}{2022}).
\newblock


\bibitem[Kang et~al\mbox{.}(2022b)]%
        {kang_from_who_you_know}
\bibfield{author}{\bibinfo{person}{Hyeonsu~B Kang}, \bibinfo{person}{Rafal
  Kocielnik}, \bibinfo{person}{Andrew Head}, \bibinfo{person}{Jiangjiang Yang},
  \bibinfo{person}{Matt Latzke}, \bibinfo{person}{Aniket Kittur},
  \bibinfo{person}{Daniel~S Weld}, \bibinfo{person}{Doug Downey}, {and}
  \bibinfo{person}{Jonathan Bragg}.} \bibinfo{year}{2022}\natexlab{b}.
\newblock \showarticletitle{From Who You Know to What You Read: Augmenting
  Scientific Recommendations with Implicit Social Networks}. In
  \bibinfo{booktitle}{\emph{Proceedings of the 2022 CHI Conference on Human
  Factors in Computing Systems}} (New Orleans, LA, USA)
  \emph{(\bibinfo{series}{CHI '22})}. \bibinfo{publisher}{Association for
  Computing Machinery}, \bibinfo{address}{New York, NY, USA}, Article
  \bibinfo{articleno}{302}, \bibinfo{numpages}{23}~pages.
\newblock
\showISBNx{9781450391573}
\urldef\tempurl%
\url{https://doi.org/10.1145/3491102.3517470}
\showDOI{\tempurl}


\bibitem[Khan et~al\mbox{.}(2020)]%
        {Khan2020DesigningAE}
\bibfield{author}{\bibinfo{person}{Taslim~Arefin Khan},
  \bibinfo{person}{Dongwook Yoon}, {and} \bibinfo{person}{Joanna McGrenere}.}
  \bibinfo{year}{2020}\natexlab{}.
\newblock \showarticletitle{Designing an Eyes-Reduced Document Skimming App for
  Situational Impairments}.
\newblock \bibinfo{journal}{\emph{Proceedings of the 2020 CHI Conference on
  Human Factors in Computing Systems}} (\bibinfo{year}{2020}).
\newblock


\bibitem[Khandwala and Guo(2018)]%
        {Khandwala2018CodemotionET}
\bibfield{author}{\bibinfo{person}{Kandarp Khandwala} {and}
  \bibinfo{person}{Philip~J. Guo}.} \bibinfo{year}{2018}\natexlab{}.
\newblock \showarticletitle{Codemotion: expanding the design space of learner
  interactions with computer programming tutorial videos}.
\newblock \bibinfo{journal}{\emph{Proceedings of the Fifth Annual ACM
  Conference on Learning at Scale}} (\bibinfo{year}{2018}).
\newblock


\bibitem[Kim et~al\mbox{.}(2014a)]%
        {Kim2014DatadrivenIT}
\bibfield{author}{\bibinfo{person}{Juho Kim}, \bibinfo{person}{Philip~J. Guo},
  \bibinfo{person}{Carrie~J. Cai}, \bibinfo{person}{Shang-Wen Li},
  \bibinfo{person}{Krzysztof~Z Gajos}, {and} \bibinfo{person}{Rob Miller}.}
  \bibinfo{year}{2014}\natexlab{a}.
\newblock \showarticletitle{Data-driven interaction techniques for improving
  navigation of educational videos}.
\newblock \bibinfo{journal}{\emph{Proceedings of the 27th annual ACM symposium
  on User interface software and technology}} (\bibinfo{year}{2014}).
\newblock


\bibitem[Kim et~al\mbox{.}(2014b)]%
        {Kim2014CrowdsourcingSI}
\bibfield{author}{\bibinfo{person}{Juho Kim}, \bibinfo{person}{Phu~Tran
  Nguyen}, \bibinfo{person}{Sarah~A. Weir}, \bibinfo{person}{Philip~J. Guo},
  \bibinfo{person}{Rob Miller}, {and} \bibinfo{person}{Krzysztof~Z Gajos}.}
  \bibinfo{year}{2014}\natexlab{b}.
\newblock \showarticletitle{Crowdsourcing step-by-step information extraction
  to enhance existing how-to videos}.
\newblock \bibinfo{journal}{\emph{Proceedings of the SIGCHI Conference on Human
  Factors in Computing Systems}} (\bibinfo{year}{2014}).
\newblock


\bibitem[King et~al\mbox{.}(2009)]%
        {King2009ScholarlyJI}
\bibfield{author}{\bibinfo{person}{Donald~W. King}, \bibinfo{person}{Carol
  Tenopir}, \bibinfo{person}{Songphan Choemprayong}, {and} \bibinfo{person}{Lei
  Wu}.} \bibinfo{year}{2009}\natexlab{}.
\newblock \showarticletitle{Scholarly journal information‐seeking and reading
  patterns of faculty at five US universities}.
\newblock \bibinfo{journal}{\emph{Learned Publishing}}  \bibinfo{volume}{22}
  (\bibinfo{year}{2009}).
\newblock


\bibitem[Kinney et~al\mbox{.}(2023)]%
        {kinney2023semantic}
\bibfield{author}{\bibinfo{person}{Rodney Kinney}, \bibinfo{person}{Chloe
  Anastasiades}, \bibinfo{person}{Russell Authur}, \bibinfo{person}{Iz
  Beltagy}, \bibinfo{person}{Jonathan Bragg}, \bibinfo{person}{Alexandra
  Buraczynski}, \bibinfo{person}{Isabel Cachola}, \bibinfo{person}{Stefan
  Candra}, \bibinfo{person}{Yoganand Chandrasekhar}, \bibinfo{person}{Arman
  Cohan}, {et~al\mbox{.}}} \bibinfo{year}{2023}\natexlab{}.
\newblock \showarticletitle{The Semantic Scholar Open Data Platform}.
\newblock \bibinfo{journal}{\emph{arXiv preprint arXiv:2301.10140}}
  (\bibinfo{year}{2023}).
\newblock


\bibitem[Kittur et~al\mbox{.}(2013)]%
        {kittur_chi13_cost_benefit}
\bibfield{author}{\bibinfo{person}{Aniket Kittur}, \bibinfo{person}{Andrew~M.
  Peters}, \bibinfo{person}{Abdigani Diriye}, \bibinfo{person}{Trupti Telang},
  {and} \bibinfo{person}{Michael~R. Bove}.} \bibinfo{year}{2013}\natexlab{}.
\newblock \showarticletitle{Costs and Benefits of Structured Information
  Foraging}. In \bibinfo{booktitle}{\emph{Proceedings of the SIGCHI Conference
  on Human Factors in Computing Systems}} (Paris, France)
  \emph{(\bibinfo{series}{CHI '13})}. \bibinfo{publisher}{Association for
  Computing Machinery}, \bibinfo{address}{New York, NY, USA},
  \bibinfo{pages}{2989–2998}.
\newblock
\showISBNx{9781450318990}
\urldef\tempurl%
\url{https://doi.org/10.1145/2470654.2481415}
\showDOI{\tempurl}


\bibitem[Krosnick(2015)]%
        {Krosnick2015VideoDocC}
\bibfield{author}{\bibinfo{person}{Rebecca Krosnick}.}
  \bibinfo{year}{2015}\natexlab{}.
\newblock \showarticletitle{VideoDoc : combining videos and lecture notes for a
  better learning experience}.
\newblock


\bibitem[Kuznetsov et~al\mbox{.}(2022)]%
        {kuznetsov_fuse}
\bibfield{author}{\bibinfo{person}{Andrew Kuznetsov},
  \bibinfo{person}{Joseph~Chee Chang}, \bibinfo{person}{Nathan Hahn},
  \bibinfo{person}{Napol Rachatasumrit}, \bibinfo{person}{Bradley Breneisen},
  \bibinfo{person}{Julina Coupland}, {and} \bibinfo{person}{Aniket Kittur}.}
  \bibinfo{year}{2022}\natexlab{}.
\newblock \showarticletitle{Fuse: In-Situ Sensemaking Support in the Browser}.
  In \bibinfo{booktitle}{\emph{Proceedings of the 35th Annual ACM Symposium on
  User Interface Software and Technology}} (Bend, OR, USA)
  \emph{(\bibinfo{series}{UIST '22})}. \bibinfo{publisher}{Association for
  Computing Machinery}, \bibinfo{address}{New York, NY, USA}, Article
  \bibinfo{articleno}{34}, \bibinfo{numpages}{15}~pages.
\newblock
\showISBNx{9781450393201}
\urldef\tempurl%
\url{https://doi.org/10.1145/3526113.3545693}
\showDOI{\tempurl}


\bibitem[Latif et~al\mbox{.}(2021)]%
        {Latif2021KoriIS}
\bibfield{author}{\bibinfo{person}{Shahid Latif}, \bibinfo{person}{Zhengzhong
  Zhou}, \bibinfo{person}{Yoon Kim}, \bibinfo{person}{Fabian Beck}, {and}
  \bibinfo{person}{Nam~Wook Kim}.} \bibinfo{year}{2021}\natexlab{}.
\newblock \showarticletitle{Kori: Interactive Synthesis of Text and Charts in
  Data Documents}.
\newblock \bibinfo{journal}{\emph{IEEE Transactions on Visualization and
  Computer Graphics}}  \bibinfo{volume}{PP} (\bibinfo{year}{2021}),
  \bibinfo{pages}{1--1}.
\newblock


\bibitem[Lee et~al\mbox{.}(2010)]%
        {Lee2010GracefullyMB}
\bibfield{author}{\bibinfo{person}{Min~Kyung Lee}, \bibinfo{person}{Sara~B.
  Kiesler}, \bibinfo{person}{Jodi Forlizzi}, \bibinfo{person}{Siddhartha~S.
  Srinivasa}, {and} \bibinfo{person}{Paul~E. Rybski}.}
  \bibinfo{year}{2010}\natexlab{}.
\newblock \showarticletitle{Gracefully mitigating breakdowns in robotic
  services}.
\newblock \bibinfo{journal}{\emph{2010 5th ACM/IEEE International Conference on
  Human-Robot Interaction (HRI)}} (\bibinfo{year}{2010}),
  \bibinfo{pages}{203--210}.
\newblock


\bibitem[Lewis et~al\mbox{.}(2020)]%
        {lewis-etal-2020-bart}
\bibfield{author}{\bibinfo{person}{Mike Lewis}, \bibinfo{person}{Yinhan Liu},
  \bibinfo{person}{Naman Goyal}, \bibinfo{person}{Marjan Ghazvininejad},
  \bibinfo{person}{Abdelrahman Mohamed}, \bibinfo{person}{Omer Levy},
  \bibinfo{person}{Veselin Stoyanov}, {and} \bibinfo{person}{Luke
  Zettlemoyer}.} \bibinfo{year}{2020}\natexlab{}.
\newblock \showarticletitle{{BART}: Denoising Sequence-to-Sequence Pre-training
  for Natural Language Generation, Translation, and Comprehension}. In
  \bibinfo{booktitle}{\emph{Proceedings of the 58th Annual Meeting of the
  Association for Computational Linguistics}}. \bibinfo{publisher}{Association
  for Computational Linguistics}, \bibinfo{address}{Online},
  \bibinfo{pages}{7871--7880}.
\newblock
\urldef\tempurl%
\url{https://doi.org/10.18653/v1/2020.acl-main.703}
\showDOI{\tempurl}


\bibitem[Liu et~al\mbox{.}(2018)]%
        {Liu2018ConceptScapeCC}
\bibfield{author}{\bibinfo{person}{Ching Liu}, \bibinfo{person}{Juho Kim},
  {and} \bibinfo{person}{Hao-Chuan Wang}.} \bibinfo{year}{2018}\natexlab{}.
\newblock \showarticletitle{ConceptScape: Collaborative Concept Mapping for
  Video Learning}.
\newblock \bibinfo{journal}{\emph{Proceedings of the 2018 CHI Conference on
  Human Factors in Computing Systems}} (\bibinfo{year}{2018}).
\newblock


\bibitem[Liu et~al\mbox{.}(2019)]%
        {liu2019unakite}
\bibfield{author}{\bibinfo{person}{Michael~Xieyang Liu}, \bibinfo{person}{Jane
  Hsieh}, \bibinfo{person}{Nathan Hahn}, \bibinfo{person}{Angelina Zhou},
  \bibinfo{person}{Emily Deng}, \bibinfo{person}{Shaun Burley},
  \bibinfo{person}{Cynthia Taylor}, \bibinfo{person}{Aniket Kittur}, {and}
  \bibinfo{person}{Brad~A Myers}.} \bibinfo{year}{2019}\natexlab{}.
\newblock \showarticletitle{Unakite: Scaffolding developers' decision-making
  using the web}. In \bibinfo{booktitle}{\emph{Proceedings of the 32nd Annual
  ACM Symposium on User Interface Software and Technology}}.
  \bibinfo{pages}{67--80}.
\newblock


\bibitem[Liu et~al\mbox{.}(2022a)]%
        {liu_crystalline}
\bibfield{author}{\bibinfo{person}{Michael~Xieyang Liu},
  \bibinfo{person}{Aniket Kittur}, {and} \bibinfo{person}{Brad~A. Myers}.}
  \bibinfo{year}{2022}\natexlab{a}.
\newblock \showarticletitle{Crystalline: Lowering the Cost for Developers to
  Collect and Organize Information for Decision Making}. In
  \bibinfo{booktitle}{\emph{Proceedings of the 2022 CHI Conference on Human
  Factors in Computing Systems}} (New Orleans, LA, USA)
  \emph{(\bibinfo{series}{CHI '22})}. \bibinfo{publisher}{Association for
  Computing Machinery}, \bibinfo{address}{New York, NY, USA}, Article
  \bibinfo{articleno}{68}, \bibinfo{numpages}{16}~pages.
\newblock
\showISBNx{9781450391573}
\urldef\tempurl%
\url{https://doi.org/10.1145/3491102.3501968}
\showDOI{\tempurl}


\bibitem[Liu et~al\mbox{.}(2022b)]%
        {liu_wigglite}
\bibfield{author}{\bibinfo{person}{Michael~Xieyang Liu},
  \bibinfo{person}{Andrew Kuznetsov}, \bibinfo{person}{Yongsung Kim},
  \bibinfo{person}{Joseph~Chee Chang}, \bibinfo{person}{Aniket Kittur}, {and}
  \bibinfo{person}{Brad~A. Myers}.} \bibinfo{year}{2022}\natexlab{b}.
\newblock \showarticletitle{Wigglite: Low-Cost Information Collection and
  Triage}. In \bibinfo{booktitle}{\emph{Proceedings of the 35th Annual ACM
  Symposium on User Interface Software and Technology}} (Bend, OR, USA)
  \emph{(\bibinfo{series}{UIST '22})}. \bibinfo{publisher}{Association for
  Computing Machinery}, \bibinfo{address}{New York, NY, USA}, Article
  \bibinfo{articleno}{32}, \bibinfo{numpages}{16}~pages.
\newblock
\showISBNx{9781450393201}
\urldef\tempurl%
\url{https://doi.org/10.1145/3526113.3545661}
\showDOI{\tempurl}


\bibitem[Lo et~al\mbox{.}(2020)]%
        {lo-etal-2020-s2orc}
\bibfield{author}{\bibinfo{person}{Kyle Lo}, \bibinfo{person}{Lucy~Lu Wang},
  \bibinfo{person}{Mark Neumann}, \bibinfo{person}{Rodney Kinney}, {and}
  \bibinfo{person}{Daniel Weld}.} \bibinfo{year}{2020}\natexlab{}.
\newblock \showarticletitle{{S}2{ORC}: The Semantic Scholar Open Research
  Corpus}. In \bibinfo{booktitle}{\emph{Proceedings of the 58th Annual Meeting
  of the Association for Computational Linguistics}}.
  \bibinfo{publisher}{Association for Computational Linguistics},
  \bibinfo{address}{Online}, \bibinfo{pages}{4969--4983}.
\newblock
\urldef\tempurl%
\url{https://doi.org/10.18653/v1/2020.acl-main.447}
\showDOI{\tempurl}


\bibitem[Luther et~al\mbox{.}(2015)]%
        {luther2015crowdlines}
\bibfield{author}{\bibinfo{person}{Kurt Luther}, \bibinfo{person}{Nathan Hahn},
  \bibinfo{person}{Steven~P Dow}, {and} \bibinfo{person}{Aniket Kittur}.}
  \bibinfo{year}{2015}\natexlab{}.
\newblock \showarticletitle{Crowdlines: Supporting synthesis of diverse
  information sources through crowdsourced outlines}. In
  \bibinfo{booktitle}{\emph{Third AAAI Conference on Human Computation and
  Crowdsourcing}}.
\newblock


\bibitem[Mackinlay et~al\mbox{.}(1995)]%
        {Mackinlay1995AnOU}
\bibfield{author}{\bibinfo{person}{Jock~D. Mackinlay}, \bibinfo{person}{Ramana
  Rao}, {and} \bibinfo{person}{Stuart~K. Card}.}
  \bibinfo{year}{1995}\natexlab{}.
\newblock \showarticletitle{An organic user interface for searching citation
  links}. In \bibinfo{booktitle}{\emph{CHI '95}}.
\newblock


\bibitem[Maliniak et~al\mbox{.}(2013)]%
        {Maliniak2013TheGC}
\bibfield{author}{\bibinfo{person}{Daniel Maliniak}, \bibinfo{person}{Ryan
  Powers}, {and} \bibinfo{person}{Barbara~F. Walter}.}
  \bibinfo{year}{2013}\natexlab{}.
\newblock \showarticletitle{The Gender Citation Gap in International
  Relations}.
\newblock \bibinfo{journal}{\emph{International Organization}}
  \bibinfo{volume}{67} (\bibinfo{year}{2013}), \bibinfo{pages}{889 -- 922}.
\newblock


\bibitem[Mayer and Moreno(1998a)]%
        {Mayer1998ACT}
\bibfield{author}{\bibinfo{person}{Richard~E. Mayer} {and}
  \bibinfo{person}{Roxana Moreno}.} \bibinfo{year}{1998}\natexlab{a}.
\newblock \showarticletitle{A Cognitive Theory of Multimedia Learning:
  Implications for Design Principles}. In \bibinfo{booktitle}{\emph{CHI 1998}}.
\newblock


\bibitem[Mayer and Moreno(1998b)]%
        {Mayer1998ASE}
\bibfield{author}{\bibinfo{person}{Richard~E. Mayer} {and}
  \bibinfo{person}{Roxana Moreno}.} \bibinfo{year}{1998}\natexlab{b}.
\newblock \showarticletitle{A Split-Attention Effect in Multimedia Learning:
  Evidence for Dual Processing Systems in Working Memory}.
\newblock \bibinfo{journal}{\emph{Journal of Educational Psychology}}
  \bibinfo{volume}{90} (\bibinfo{year}{1998}), \bibinfo{pages}{312--320}.
\newblock


\bibitem[McKiernan et~al\mbox{.}(2016)]%
        {mckiernan2016open}
\bibfield{author}{\bibinfo{person}{Erin~C McKiernan}, \bibinfo{person}{Philip~E
  Bourne}, \bibinfo{person}{C~Titus Brown}, \bibinfo{person}{Stuart Buck},
  \bibinfo{person}{Amye Kenall}, \bibinfo{person}{Jennifer Lin},
  \bibinfo{person}{Damon McDougall}, \bibinfo{person}{Brian~A Nosek},
  \bibinfo{person}{Karthik Ram}, \bibinfo{person}{Courtney~K Soderberg},
  {et~al\mbox{.}}} \bibinfo{year}{2016}\natexlab{}.
\newblock \showarticletitle{How open science helps researchers succeed}.
\newblock \bibinfo{journal}{\emph{elife}}  \bibinfo{volume}{5}
  (\bibinfo{year}{2016}).
\newblock


\bibitem[McKiernan(2000)]%
        {mckiernan2000arxiv}
\bibfield{author}{\bibinfo{person}{Gerry McKiernan}.}
  \bibinfo{year}{2000}\natexlab{}.
\newblock \showarticletitle{arXiv.org: the Los Alamos National Laboratory
  e-print server}.
\newblock \bibinfo{journal}{\emph{International Journal on Grey Literature}}
  (\bibinfo{year}{2000}).
\newblock


\bibitem[Murthy et~al\mbox{.}(2022)]%
        {murthy2022accord}
\bibfield{author}{\bibinfo{person}{Sonia~K Murthy}, \bibinfo{person}{Kyle Lo},
  \bibinfo{person}{Daniel King}, \bibinfo{person}{Chandra Bhagavatula},
  \bibinfo{person}{Bailey Kuehl}, \bibinfo{person}{Sophie Johnson},
  \bibinfo{person}{Jonathan Borchardt}, \bibinfo{person}{Daniel~S Weld},
  \bibinfo{person}{Tom Hope}, {and} \bibinfo{person}{Doug Downey}.}
  \bibinfo{year}{2022}\natexlab{}.
\newblock \showarticletitle{ACCoRD: A Multi-Document Approach to Generating
  Diverse Descriptions of Scientific Concepts}.
\newblock \bibinfo{journal}{\emph{arXiv preprint arXiv:2205.06982}}
  (\bibinfo{year}{2022}).
\newblock


\bibitem[Nakov et~al\mbox{.}(2004)]%
        {nakov2004citances}
\bibfield{author}{\bibinfo{person}{Preslav~I Nakov}, \bibinfo{person}{Ariel~S
  Schwartz}, \bibinfo{person}{Marti Hearst}, {et~al\mbox{.}}}
  \bibinfo{year}{2004}\natexlab{}.
\newblock \showarticletitle{Citances: Citation sentences for semantic analysis
  of bioscience text}. In \bibinfo{booktitle}{\emph{Proceedings of the SIGIR}},
  Vol.~\bibinfo{volume}{4}. Citeseer, \bibinfo{pages}{81--88}.
\newblock


\bibitem[Okamura(2019)]%
        {okamura2019interdisciplinarity}
\bibfield{author}{\bibinfo{person}{Keisuke Okamura}.}
  \bibinfo{year}{2019}\natexlab{}.
\newblock \showarticletitle{Interdisciplinarity revisited: evidence for
  research impact and dynamism}.
\newblock \bibinfo{journal}{\emph{Palgrave Communications}}
  \bibinfo{volume}{5}, \bibinfo{number}{1} (\bibinfo{year}{2019}),
  \bibinfo{pages}{1--9}.
\newblock


\bibitem[Otmakhova et~al\mbox{.}(2022)]%
        {otmakhova-etal-2022-patient}
\bibfield{author}{\bibinfo{person}{Yulia Otmakhova}, \bibinfo{person}{Karin
  Verspoor}, \bibinfo{person}{Timothy Baldwin}, {and} \bibinfo{person}{Jey~Han
  Lau}.} \bibinfo{year}{2022}\natexlab{}.
\newblock \showarticletitle{The patient is more dead than alive: exploring the
  current state of the multi-document summarisation of the biomedical
  literature}. In \bibinfo{booktitle}{\emph{Proceedings of the 60th Annual
  Meeting of the Association for Computational Linguistics (Volume 1: Long
  Papers)}}. \bibinfo{publisher}{Association for Computational Linguistics},
  \bibinfo{address}{Dublin, Ireland}, \bibinfo{pages}{5098--5111}.
\newblock
\urldef\tempurl%
\url{https://doi.org/10.18653/v1/2022.acl-long.350}
\showDOI{\tempurl}


\bibitem[Palani et~al\mbox{.}(2023)]%
        {relatedly}
\bibfield{author}{\bibinfo{person}{Srishti Palani}, \bibinfo{person}{Aakanksha
  Naik}, \bibinfo{person}{Doug Downey}, \bibinfo{person}{Amy~X. Zhang},
  \bibinfo{person}{Jonathan Bragg}, {and} \bibinfo{person}{Joseph~Chee Chang}.}
  \bibinfo{year}{2023}\natexlab{}.
\newblock \showarticletitle{{Relatedly}: Scaffolding Literature Reviews with
  Existing Related Work Sections}.
\newblock   \bibinfo{volume}{22} (\bibinfo{year}{2023}).
\newblock


\bibitem[Palmer et~al\mbox{.}(2009)]%
        {palmer2009scholarly}
\bibfield{author}{\bibinfo{person}{Carole~L Palmer}, \bibinfo{person}{Lauren~C
  Teffeau}, {and} \bibinfo{person}{Carrie~M Pirmann}.}
  \bibinfo{year}{2009}\natexlab{}.
\newblock \showarticletitle{Scholarly information practices in the online
  environment}.
\newblock \bibinfo{journal}{\emph{Report commissioned by OCLC Research.
  Published online at: www. oclc. org/programs/publications/reports/2009-02.
  pdf}} (\bibinfo{year}{2009}).
\newblock


\bibitem[Park et~al\mbox{.}(2022)]%
        {park-cscw22}
\bibfield{author}{\bibinfo{person}{Soya Park}, \bibinfo{person}{Jonathan
  Bragg}, \bibinfo{person}{Michael Chang}, \bibinfo{person}{Kevin Larson},
  {and} \bibinfo{person}{Danielle Bragg}.} \bibinfo{year}{2022}\natexlab{}.
\newblock \showarticletitle{Exploring Team-Sourced Hyperlinks to Address
  Navigation Challenges for Low-Vision Readers of Scientific Papers}. In
  \bibinfo{booktitle}{\emph{Proceedings of the 25th ACM Conference On
  Computer-Supported Cooperative Work And Social Computing}}.
\newblock


\bibitem[Pavel et~al\mbox{.}(2014)]%
        {Pavel2014VideoDA}
\bibfield{author}{\bibinfo{person}{Amy Pavel}, \bibinfo{person}{Colorado Reed},
  \bibinfo{person}{Bj{\"o}rn Hartmann}, {and} \bibinfo{person}{Maneesh
  Agrawala}.} \bibinfo{year}{2014}\natexlab{}.
\newblock \showarticletitle{Video digests: a browsable, skimmable format for
  informational lecture videos}.
\newblock \bibinfo{journal}{\emph{Proceedings of the 27th annual ACM symposium
  on User interface software and technology}} (\bibinfo{year}{2014}).
\newblock


\bibitem[Peroni et~al\mbox{.}(2015)]%
        {Peroni2015SettingOB}
\bibfield{author}{\bibinfo{person}{Silvio Peroni}, \bibinfo{person}{Alexander
  Dutton}, \bibinfo{person}{Tanya Gray}, {and} \bibinfo{person}{David~M.
  Shotton}.} \bibinfo{year}{2015}\natexlab{}.
\newblock \showarticletitle{Setting our bibliographic references free: towards
  open citation data}.
\newblock \bibinfo{journal}{\emph{J. Documentation}}  \bibinfo{volume}{71}
  (\bibinfo{year}{2015}), \bibinfo{pages}{253--277}.
\newblock


\bibitem[Philip et~al\mbox{.}(2014)]%
        {Philip2014ApplicationOC}
\bibfield{author}{\bibinfo{person}{Simon Philip},
  \bibinfo{person}{Peter~Bamidele Shola}, {and} \bibinfo{person}{Abari~Ovye
  John}.} \bibinfo{year}{2014}\natexlab{}.
\newblock \showarticletitle{Application of Content-Based Approach in Research
  Paper Recommendation System for a Digital Library}.
\newblock \bibinfo{journal}{\emph{International Journal of Advanced Computer
  Science and Applications}}  \bibinfo{volume}{5} (\bibinfo{year}{2014}).
\newblock


\bibitem[Pirolli and Card(1999)]%
        {Pirolli_InformationForaging_1999}
\bibfield{author}{\bibinfo{person}{Peter Pirolli} {and} \bibinfo{person}{Stuart
  Card}.} \bibinfo{year}{1999}\natexlab{}.
\newblock \showarticletitle{Information Foraging}.
\newblock \bibinfo{journal}{\emph{Psychological Review}} \bibinfo{volume}{106},
  \bibinfo{number}{4} (\bibinfo{year}{1999}), \bibinfo{pages}{643--675}.
\newblock
\urldef\tempurl%
\url{https://doi.org/10.1037/0033-295x.106.4.643}
\showDOI{\tempurl}


\bibitem[Ponsard et~al\mbox{.}(2016)]%
        {Ponsard2016PaperQuestAV}
\bibfield{author}{\bibinfo{person}{Antoine Ponsard}, \bibinfo{person}{Francisco
  Escalona}, {and} \bibinfo{person}{Tamara Munzner}.}
  \bibinfo{year}{2016}\natexlab{}.
\newblock \showarticletitle{PaperQuest: A Visualization Tool to Support
  Literature Review}.
\newblock \bibinfo{journal}{\emph{Proceedings of the 2016 CHI Conference
  Extended Abstracts on Human Factors in Computing Systems}}
  (\bibinfo{year}{2016}).
\newblock


\bibitem[Portenoy et~al\mbox{.}(2022)]%
        {portenoy2022bursting}
\bibfield{author}{\bibinfo{person}{Jason Portenoy}, \bibinfo{person}{Marissa
  Radensky}, \bibinfo{person}{Jevin~D West}, \bibinfo{person}{Eric Horvitz},
  \bibinfo{person}{Daniel~S Weld}, {and} \bibinfo{person}{Tom Hope}.}
  \bibinfo{year}{2022}\natexlab{}.
\newblock \showarticletitle{Bursting scientific filter bubbles: Boosting
  innovation via novel author discovery}. In
  \bibinfo{booktitle}{\emph{Proceedings of the 2022 CHI Conference on Human
  Factors in Computing Systems}}. \bibinfo{pages}{1--13}.
\newblock


\bibitem[Rachatasumrit et~al\mbox{.}(2022)]%
        {Rachatasumrit2022CiteReadIL}
\bibfield{author}{\bibinfo{person}{Napol Rachatasumrit},
  \bibinfo{person}{Jonathan Bragg}, \bibinfo{person}{Amy~X. Zhang}, {and}
  \bibinfo{person}{Daniel~S. Weld}.} \bibinfo{year}{2022}\natexlab{}.
\newblock \showarticletitle{CiteRead: Integrating Localized Citation Contexts
  into Scientific Paper Reading}.
\newblock \bibinfo{journal}{\emph{27th International Conference on Intelligent
  User Interfaces}} (\bibinfo{year}{2022}).
\newblock


\bibitem[Raffel et~al\mbox{.}(2019)]%
        {Raffel2019ExploringTL}
\bibfield{author}{\bibinfo{person}{Colin Raffel}, \bibinfo{person}{Noam~M.
  Shazeer}, \bibinfo{person}{Adam Roberts}, \bibinfo{person}{Katherine Lee},
  \bibinfo{person}{Sharan Narang}, \bibinfo{person}{Michael Matena},
  \bibinfo{person}{Yanqi Zhou}, \bibinfo{person}{Wei Li}, {and}
  \bibinfo{person}{Peter~J. Liu}.} \bibinfo{year}{2019}\natexlab{}.
\newblock \showarticletitle{Exploring the Limits of Transfer Learning with a
  Unified Text-to-Text Transformer}.
\newblock \bibinfo{journal}{\emph{ArXiv}}  \bibinfo{volume}{abs/1910.10683}
  (\bibinfo{year}{2019}).
\newblock


\bibitem[Russell et~al\mbox{.}(1993)]%
        {Russell_Sensemaking_1993}
\bibfield{author}{\bibinfo{person}{Daniel~M. Russell}, \bibinfo{person}{Mark~J.
  Stefik}, \bibinfo{person}{Peter Pirolli}, {and} \bibinfo{person}{Stuart~K.
  Card}.} \bibinfo{year}{1993}\natexlab{}.
\newblock \showarticletitle{The Cost Structure of Sensemaking}. In
  \bibinfo{booktitle}{\emph{Proceedings of the INTERACT '93 and CHI '93
  Conference on Human Factors in Computing Systems}} (Amsterdam, The
  Netherlands) \emph{(\bibinfo{series}{CHI '93})}.
  \bibinfo{publisher}{Association for Computing Machinery},
  \bibinfo{address}{New York, NY, USA}, \bibinfo{pages}{269–276}.
\newblock
\showISBNx{0897915755}
\urldef\tempurl%
\url{https://doi.org/10.1145/169059.169209}
\showDOI{\tempurl}


\bibitem[Shahaf et~al\mbox{.}(2012)]%
        {shahaf2012metro}
\bibfield{author}{\bibinfo{person}{Dafna Shahaf}, \bibinfo{person}{Carlos
  Guestrin}, {and} \bibinfo{person}{Eric Horvitz}.}
  \bibinfo{year}{2012}\natexlab{}.
\newblock \showarticletitle{Metro maps of science}. In
  \bibinfo{booktitle}{\emph{Proceedings of the 18th ACM SIGKDD international
  conference on Knowledge discovery and data mining}}.
  \bibinfo{pages}{1122--1130}.
\newblock


\bibitem[Shen et~al\mbox{.}(2022)]%
        {shen-etal-2022-vila}
\bibfield{author}{\bibinfo{person}{Zejiang Shen}, \bibinfo{person}{Kyle Lo},
  \bibinfo{person}{Lucy~Lu Wang}, \bibinfo{person}{Bailey Kuehl},
  \bibinfo{person}{Daniel~S. Weld}, {and} \bibinfo{person}{Doug Downey}.}
  \bibinfo{year}{2022}\natexlab{}.
\newblock \showarticletitle{{VILA}: Improving Structured Content Extraction
  from Scientific {PDF}s Using Visual Layout Groups}.
\newblock \bibinfo{journal}{\emph{Transactions of the Association for
  Computational Linguistics}}  \bibinfo{volume}{10} (\bibinfo{year}{2022}),
  \bibinfo{pages}{376--392}.
\newblock
\urldef\tempurl%
\url{https://doi.org/10.1162/tacl_a_00466}
\showDOI{\tempurl}


\bibitem[Shneiderman(2022)]%
        {shneiderman2022human}
\bibfield{author}{\bibinfo{person}{Ben Shneiderman}.}
  \bibinfo{year}{2022}\natexlab{}.
\newblock \bibinfo{booktitle}{\emph{Human-centered AI}}.
\newblock \bibinfo{publisher}{Oxford University Press}.
\newblock


\bibitem[Subramonyam et~al\mbox{.}(2020)]%
        {texSketch}
\bibfield{author}{\bibinfo{person}{Hariharan Subramonyam},
  \bibinfo{person}{Colleen Seifert}, \bibinfo{person}{Priti Shah}, {and}
  \bibinfo{person}{Eytan Adar}.} \bibinfo{year}{2020}\natexlab{}.
\newblock \showarticletitle{TexSketch: Active Diagramming through Pen-and-Ink
  Annotations}. In \bibinfo{booktitle}{\emph{Proceedings of the 2020 CHI
  Conference on Human Factors in Computing Systems}} (Honolulu, HI, USA)
  \emph{(\bibinfo{series}{CHI '20})}. \bibinfo{publisher}{Association for
  Computing Machinery}, \bibinfo{address}{New York, NY, USA},
  \bibinfo{pages}{1–13}.
\newblock
\showISBNx{9781450367080}
\urldef\tempurl%
\url{https://doi.org/10.1145/3313831.3376155}
\showDOI{\tempurl}


\bibitem[Sugiyama and Kan(2010)]%
        {Sugiyama2010ScholarlyPR}
\bibfield{author}{\bibinfo{person}{Kazunari Sugiyama} {and}
  \bibinfo{person}{Min-Yen Kan}.} \bibinfo{year}{2010}\natexlab{}.
\newblock \showarticletitle{Scholarly paper recommendation via user's recent
  research interests}. In \bibinfo{booktitle}{\emph{JCDL '10}}.
\newblock


\bibitem[Szpiro et~al\mbox{.}(2016)]%
        {Szpiro2016HowPW}
\bibfield{author}{\bibinfo{person}{Sarit Felicia~Anais Szpiro},
  \bibinfo{person}{Shafeka Hashash}, \bibinfo{person}{Yuhang Zhao}, {and}
  \bibinfo{person}{Shiri Azenkot}.} \bibinfo{year}{2016}\natexlab{}.
\newblock \showarticletitle{How People with Low Vision Access Computing
  Devices: Understanding Challenges and Opportunities}.
\newblock \bibinfo{journal}{\emph{Proceedings of the 18th International ACM
  SIGACCESS Conference on Computers and Accessibility}} (\bibinfo{year}{2016}).
\newblock


\bibitem[team(tion)]%
        {paper2html}
\bibfield{author}{\bibinfo{person}{Semantic~Reader team}.}
  \bibinfo{year}{Unpublished demo application}\natexlab{}.
\newblock \bibinfo{title}{Paper to HTML}.
\newblock
\newblock
\urldef\tempurl%
\url{https://papertohtml.org}
\showURL{%
\tempurl}


\bibitem[team(sion)]%
        {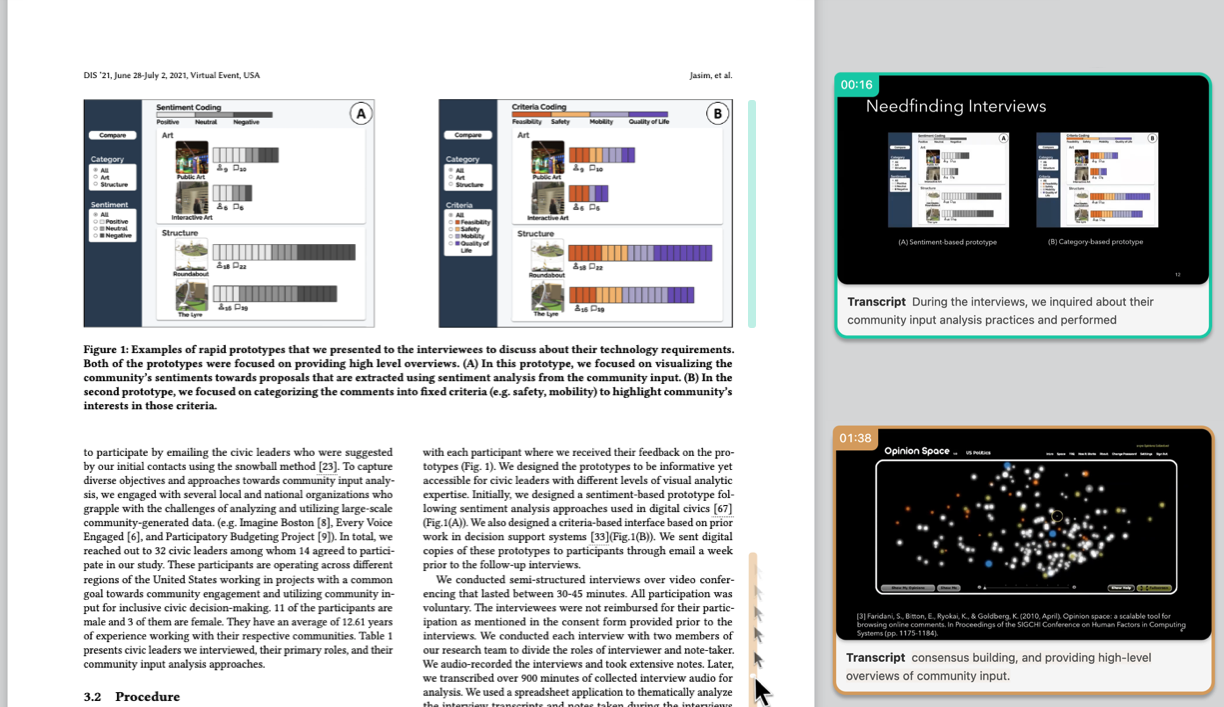}
\bibfield{author}{\bibinfo{person}{Semantic~Reader team}.}
  \bibinfo{year}{Unpublished demo application; in submission.}\natexlab{}.
\newblock \bibinfo{title}{Papeo}.
\newblock
\newblock
\urldef\tempurl%
\url{https://papeo.app}
\showURL{%
\tempurl}


\bibitem[Truong et~al\mbox{.}(2021)]%
        {Truong2021AutomaticGO}
\bibfield{author}{\bibinfo{person}{Anh~Tuan Truong}, \bibinfo{person}{Peggy
  Chi}, \bibinfo{person}{D. Salesin}, \bibinfo{person}{Irfan Essa}, {and}
  \bibinfo{person}{Maneesh Agrawala}.} \bibinfo{year}{2021}\natexlab{}.
\newblock \showarticletitle{Automatic Generation of Two-Level Hierarchical
  Tutorials from Instructional Makeup Videos}.
\newblock \bibinfo{journal}{\emph{Proceedings of the 2021 CHI Conference on
  Human Factors in Computing Systems}} (\bibinfo{year}{2021}).
\newblock


\bibitem[Van~Noorden et~al\mbox{.}(2015)]%
        {van2015interdisciplinary}
\bibfield{author}{\bibinfo{person}{Richard Van~Noorden} {et~al\mbox{.}}}
  \bibinfo{year}{2015}\natexlab{}.
\newblock \showarticletitle{Interdisciplinary research by the numbers}.
\newblock \bibinfo{journal}{\emph{Nature}} \bibinfo{volume}{525},
  \bibinfo{number}{7569} (\bibinfo{year}{2015}), \bibinfo{pages}{306--307}.
\newblock


\bibitem[Wadden et~al\mbox{.}(2020)]%
        {wadden-etal-2020-fact}
\bibfield{author}{\bibinfo{person}{David Wadden}, \bibinfo{person}{Shanchuan
  Lin}, \bibinfo{person}{Kyle Lo}, \bibinfo{person}{Lucy~Lu Wang},
  \bibinfo{person}{Madeleine van Zuylen}, \bibinfo{person}{Arman Cohan}, {and}
  \bibinfo{person}{Hannaneh Hajishirzi}.} \bibinfo{year}{2020}\natexlab{}.
\newblock \showarticletitle{Fact or Fiction: Verifying Scientific Claims}. In
  \bibinfo{booktitle}{\emph{Proceedings of the 2020 Conference on Empirical
  Methods in Natural Language Processing (EMNLP)}}.
  \bibinfo{publisher}{Association for Computational Linguistics},
  \bibinfo{address}{Online}, \bibinfo{pages}{7534--7550}.
\newblock
\urldef\tempurl%
\url{https://doi.org/10.18653/v1/2020.emnlp-main.609}
\showDOI{\tempurl}


\bibitem[Wang and Blei(2011)]%
        {Wang2011CollaborativeTM}
\bibfield{author}{\bibinfo{person}{Chong Wang} {and} \bibinfo{person}{David~M.
  Blei}.} \bibinfo{year}{2011}\natexlab{}.
\newblock \showarticletitle{Collaborative topic modeling for recommending
  scientific articles}. In \bibinfo{booktitle}{\emph{KDD}}.
\newblock


\bibitem[Wang et~al\mbox{.}(2021b)]%
        {wang-2021-scia11y}
\bibfield{author}{\bibinfo{person}{Lucy~Lu Wang}, \bibinfo{person}{Isabel
  Cachola}, \bibinfo{person}{Jonathan Bragg}, \bibinfo{person}{Evie Yu-Yen
  Cheng}, \bibinfo{person}{Chelsea Haupt}, \bibinfo{person}{Matt Latzke},
  \bibinfo{person}{Bailey Kuehl}, \bibinfo{person}{Madeleine~N van Zuylen},
  \bibinfo{person}{Linda Wagner}, {and} \bibinfo{person}{Daniel Weld}.}
  \bibinfo{year}{2021}\natexlab{b}.
\newblock \showarticletitle{SciA11y: Converting Scientific Papers to Accessible
  HTML}. In \bibinfo{booktitle}{\emph{Proceedings of the 23rd International ACM
  SIGACCESS Conference on Computers and Accessibility}} (Virtual Event, USA)
  \emph{(\bibinfo{series}{ASSETS '21})}. \bibinfo{publisher}{Association for
  Computing Machinery}, \bibinfo{address}{New York, NY, USA}, Article
  \bibinfo{articleno}{85}, \bibinfo{numpages}{4}~pages.
\newblock
\showISBNx{9781450383066}
\urldef\tempurl%
\url{https://doi.org/10.1145/3441852.3476545}
\showDOI{\tempurl}


\bibitem[Wang et~al\mbox{.}(2021c)]%
        {wang-2021-accessibility}
\bibfield{author}{\bibinfo{person}{Lucy~Lu Wang}, \bibinfo{person}{Isabel
  Cachola}, \bibinfo{person}{Jonathan Bragg}, \bibinfo{person}{Evie Yu-Yen
  Cheng}, \bibinfo{person}{Chelsea~Hess Haupt}, \bibinfo{person}{Matt Latzke},
  \bibinfo{person}{Bailey Kuehl}, \bibinfo{person}{Madeleine {van Zuylen}},
  \bibinfo{person}{Linda Wagner}, {and} \bibinfo{person}{Daniel~S. Weld}.}
  \bibinfo{year}{2021}\natexlab{c}.
\newblock \showarticletitle{Improving the {{Accessibility}} of {{Scientific
  Documents}}: {{Current State}}, {{User Needs}}, and a {{System Solution}} to
  {{Enhance Scientific PDF Accessibility}} for {{Blind}} and {{Low Vision
  Users}}}.
\newblock \bibinfo{journal}{\emph{arXiv: 2105.00076 [cs.DL]}}
  (\bibinfo{year}{2021}).
\newblock
\showeprint[arxiv]{2105.00076}~[cs.DL]


\bibitem[Wang et~al\mbox{.}(2021a)]%
        {wang-etal-2021-minilmv2}
\bibfield{author}{\bibinfo{person}{Wenhui Wang}, \bibinfo{person}{Hangbo Bao},
  \bibinfo{person}{Shaohan Huang}, \bibinfo{person}{Li Dong}, {and}
  \bibinfo{person}{Furu Wei}.} \bibinfo{year}{2021}\natexlab{a}.
\newblock \showarticletitle{{M}ini{LM}v2: Multi-Head Self-Attention Relation
  Distillation for Compressing Pretrained Transformers}. In
  \bibinfo{booktitle}{\emph{Findings of the Association for Computational
  Linguistics: ACL-IJCNLP 2021}}. \bibinfo{publisher}{Association for
  Computational Linguistics}, \bibinfo{address}{Online},
  \bibinfo{pages}{2140--2151}.
\newblock
\urldef\tempurl%
\url{https://doi.org/10.18653/v1/2021.findings-acl.188}
\showDOI{\tempurl}


\bibitem[Wang et~al\mbox{.}(2020)]%
        {wang-2020-minilm-v1}
\bibfield{author}{\bibinfo{person}{Wenhui Wang}, \bibinfo{person}{Furu Wei},
  \bibinfo{person}{Li Dong}, \bibinfo{person}{Hangbo Bao}, \bibinfo{person}{Nan
  Yang}, {and} \bibinfo{person}{Ming Zhou}.} \bibinfo{year}{2020}\natexlab{}.
\newblock \showarticletitle{MiniLM: Deep Self-Attention Distillation for
  Task-Agnostic Compression of Pre-Trained Transformers}. In
  \bibinfo{booktitle}{\emph{Advances in Neural Information Processing
  Systems}}, \bibfield{editor}{\bibinfo{person}{H.~Larochelle},
  \bibinfo{person}{M.~Ranzato}, \bibinfo{person}{R.~Hadsell},
  \bibinfo{person}{M.F. Balcan}, {and} \bibinfo{person}{H.~Lin}} (Eds.),
  Vol.~\bibinfo{volume}{33}. \bibinfo{publisher}{Curran Associates, Inc.},
  \bibinfo{pages}{5776--5788}.
\newblock
\urldef\tempurl%
\url{https://proceedings.neurips.cc/paper_files/paper/2020/file/3f5ee243547dee91fbd053c1c4a845aa-Paper.pdf}
\showURL{%
\tempurl}


\bibitem[Way et~al\mbox{.}(2019)]%
        {way2019productivity}
\bibfield{author}{\bibinfo{person}{Samuel~F Way}, \bibinfo{person}{Allison~C
  Morgan}, \bibinfo{person}{Daniel~B Larremore}, {and} \bibinfo{person}{Aaron
  Clauset}.} \bibinfo{year}{2019}\natexlab{}.
\newblock \showarticletitle{Productivity, prominence, and the effects of
  academic environment}.
\newblock \bibinfo{journal}{\emph{Proceedings of the National Academy of
  Sciences}} \bibinfo{volume}{116}, \bibinfo{number}{22}
  (\bibinfo{year}{2019}), \bibinfo{pages}{10729--10733}.
\newblock


\bibitem[Wecker et~al\mbox{.}(2014)]%
        {wecker_semantize_2014}
\bibfield{author}{\bibinfo{person}{Alan~J. Wecker}, \bibinfo{person}{Joel
  Lanir}, \bibinfo{person}{Osnat Mokryn}, \bibinfo{person}{Einat Minkov}, {and}
  \bibinfo{person}{Tsvi Kuflik}.} \bibinfo{year}{2014}\natexlab{}.
\newblock \showarticletitle{Semantize: visualizing the sentiment of individual
  document}. In \bibinfo{booktitle}{\emph{Proceedings of the 2014 International
  Working Conference on Advanced Visual Interfaces}}.
  \bibinfo{publisher}{Association for Computing Machinery},
  \bibinfo{address}{Como, Italy}, \bibinfo{pages}{385--386}.
\newblock


\bibitem[Xia et~al\mbox{.}(2016)]%
        {Xia2016ScientificAR}
\bibfield{author}{\bibinfo{person}{Feng Xia}, \bibinfo{person}{Haifeng Liu},
  \bibinfo{person}{Ivan Lee}, {and} \bibinfo{person}{Longbing Cao}.}
  \bibinfo{year}{2016}\natexlab{}.
\newblock \showarticletitle{Scientific Article Recommendation: Exploiting
  Common Author Relations and Historical Preferences}.
\newblock \bibinfo{journal}{\emph{IEEE Transactions on Big Data}}
  \bibinfo{volume}{2} (\bibinfo{year}{2016}), \bibinfo{pages}{101--G112}.
\newblock


\bibitem[Xu et~al\mbox{.}(2019)]%
        {Xu2019LayoutLMPO}
\bibfield{author}{\bibinfo{person}{Yiheng Xu}, \bibinfo{person}{Minghao Li},
  \bibinfo{person}{Lei Cui}, \bibinfo{person}{Shaohan Huang},
  \bibinfo{person}{Furu Wei}, {and} \bibinfo{person}{Ming Zhou}.}
  \bibinfo{year}{2019}\natexlab{}.
\newblock \showarticletitle{LayoutLM: Pre-training of Text and Layout for
  Document Image Understanding}.
\newblock \bibinfo{journal}{\emph{Proceedings of the 26th ACM SIGKDD
  International Conference on Knowledge Discovery \& Data Mining}}
  (\bibinfo{year}{2019}).
\newblock


\bibitem[Yang et~al\mbox{.}(2017)]%
        {yang_hitext_2017}
\bibfield{author}{\bibinfo{person}{Qian Yang}, \bibinfo{person}{Gerard de
  Melo}, \bibinfo{person}{Yong Cheng}, {and} \bibinfo{person}{Sen Wang}.}
  \bibinfo{year}{2017}\natexlab{}.
\newblock \showarticletitle{HiText: Text Reading with Dynamic Salience
  Marking}. In \bibinfo{booktitle}{\emph{Proceedings of the 26th International
  Conference on World Wide Web Companion}}. \bibinfo{publisher}{Association for
  Computing Machinery}, \bibinfo{address}{Perth, Australia},
  \bibinfo{pages}{311--319}.
\newblock


\bibitem[Yoon et~al\mbox{.}(2019)]%
        {Yoon2019PretrainedLM}
\bibfield{author}{\bibinfo{person}{Wonjin Yoon}, \bibinfo{person}{Jinhyuk Lee},
  \bibinfo{person}{Donghyeon Kim}, \bibinfo{person}{Minbyul Jeong}, {and}
  \bibinfo{person}{Jaewoo Kang}.} \bibinfo{year}{2019}\natexlab{}.
\newblock \showarticletitle{Pre-trained Language Model for Biomedical Question
  Answering}. In \bibinfo{booktitle}{\emph{PKDD/ECML Workshops}}.
\newblock


\bibitem[Zellweger et~al\mbox{.}(1998)]%
        {ref:zellweger1998fluid}
\bibfield{author}{\bibinfo{person}{Polle~T. Zellweger},
  \bibinfo{person}{Bay-Wei Chang}, {and} \bibinfo{person}{Jock~D. Mackinlay}.}
  \bibinfo{year}{1998}\natexlab{}.
\newblock \showarticletitle{Fluid Links for Informed and Incremental Link
  Transitions}. In \bibinfo{booktitle}{\emph{Proceedings of the Conference on
  Hypertext and Hypermedia}}. \bibinfo{publisher}{ACM},
  \bibinfo{pages}{50--57}.
\newblock


\bibitem[Zhao and Lee(2020)]%
        {Zhao2020TalkTP}
\bibfield{author}{\bibinfo{person}{Tian Zhao} {and} \bibinfo{person}{Kyusong
  Lee}.} \bibinfo{year}{2020}\natexlab{}.
\newblock \showarticletitle{Talk to Papers: Bringing Neural Question Answering
  to Academic Search}.
\newblock \bibinfo{journal}{\emph{ArXiv}}  \bibinfo{volume}{abs/2004.02002}
  (\bibinfo{year}{2020}).
\newblock


\bibitem[Zyto et~al\mbox{.}(2012)]%
        {zyto2012successful}
\bibfield{author}{\bibinfo{person}{Sacha Zyto}, \bibinfo{person}{David Karger},
  \bibinfo{person}{Mark Ackerman}, {and} \bibinfo{person}{Sanjoy Mahajan}.}
  \bibinfo{year}{2012}\natexlab{}.
\newblock \showarticletitle{Successful classroom deployment of a social
  document annotation system}. In \bibinfo{booktitle}{\emph{Proceedings of the
  sigchi conference on human factors in computing systems}}.
  \bibinfo{pages}{1883--1892}.
\newblock


\end{thebibliography}

\appendix

\balance

\end{document}